\documentclass[journal, onecolumn]{aastex631}
\usepackage{amsmath}
\usepackage{CJK}
\usepackage{booktabs}
\usepackage{threeparttable}
\usepackage{amsthm,amsmath,amssymb}
\usepackage{mathrsfs}

\shorttitle{xing et al}
\shortauthors{xing et al}
\graphicspath{{./}{figures/}}

\begin{document}
\begin{CJK*}{UTF8}{gbsn}

\title{The Mass Fractionation of Helium in the Escaping Atmosphere of HD 209458b\footnote{Revised Manuscript on June, 7, 2022}}

\email{guojh@ynao.ac.cn}

\author[0000-0001-7655-0920]{lei xing (邢磊)}
\affiliation{Yunnan Observatories, Chinese Academy of Sciences \\
P.0.Box110 \\
Kunming, 650216, People’s Republic of China}
\affiliation{University of Chinese Academy of Sciences \\
No.19(A) Yuquan Road, Shijingshan District \\
Beijing, 100049, People’s Republic of China}
\affiliation{Key Laboratory for Structure and Evolution of Celestial Objects, Chinese Academy of Sciences \\
P.0.Box110 \\
Kunming, 650216, People’s Republic of China}

\author[0000-0002-8519-0514]{dongdong yan (闫冬冬)}
\affiliation{Yunnan Observatories, Chinese Academy of Sciences \\
P.0.Box110 \\
Kunming, 650216, People’s Republic of China}
\affiliation{University of Chinese Academy of Sciences \\
No.19(A) Yuquan Road, Shijingshan District \\
Beijing, 100049, People’s Republic of China}
\affiliation{Key Laboratory for Structure and Evolution of Celestial Objects, Chinese Academy of Sciences \\
P.0.Box110 \\
Kunming, 650216, People’s Republic of China}

\author[0000-0002-8869-6510]{jianheng guo (郭建恒)}
\affiliation{Yunnan Observatories, Chinese Academy of Sciences \\
P.0.Box110 \\
Kunming, 650216, People’s Republic of China}
\affiliation{University of Chinese Academy of Sciences \\
No.19(A) Yuquan Road, Shijingshan District \\
Beijing, 100049, People’s Republic of China}
\affiliation{Key Laboratory for Structure and Evolution of Celestial Objects, Chinese Academy of Sciences \\
P.0.Box110 \\
Kunming, 650216, People’s Republic of China}


%
%
%



\begin{abstract}
The absorption signals of metastable He in HD 209458b and several other exoplanets can be explained via escaping atmosphere model with a subsolar He/H ratio. The low abundance of helium can be a result of planet formation if there is a small amount of helium in their primordial atmosphere. However, another possibility is that the low He/H ratio is caused by the process of mass fractionation of helium in the atmosphere. In order to investigate the effect of the fractionation in the hydrogen-helium atmosphere, we developed a self-consistent multi-fluid 1D hydrodynamic model based on the well-known open-source MHD code PLUTO. Our simulations show that a lower He/H ratio can be produced spontaneously in the multi-fluid model. We further modeled the transmission spectra of He 10830 lines for HD 209458b in a broad parameter space. The transmission spectrum of the observation can be fitted in the condition of 1.80 times the X-ray and extreme-ultraviolet flux of the quiet Sun. Meanwhile, the ratio of the escaping flux of helium to hydrogen, $F_{He}/F_{H}$, is 0.039. Our results indicate that the mass fractionation of helium to hydrogen can naturally interpret the low He/H ratio required by the observation. Thus, in the escaping atmosphere of HD 209458b, decreasing the abundance of helium in the atmosphere is not needed even if its He abundance is similar to that of the Sun. The simulation presented in this work hints that in the escaping atmosphere, mass fractionation can also occur on other exoplanets, which needs to be explored further.
\end{abstract}



\section{Introduction} \label{sec:intro}

Under strong stellar X-ray and extreme-ultraviolet (XUV) irradiation, close-in exoplanets can experience hydrodynamic escape in their upper atmosphere, which is several orders of magnitude larger than the well-known Jeans escape \citep{Murray2009}. The massive atmosphere escape has a great influence on planetary evolution \citep{Lecavelier2004}. For instance, \cite{Owen2017} suggest that the dearth of short-period Neptunian exoplanets \citep{Mazeh2016} is a result of fast mass loss of atmosphere. A prerequisite for estimating the process of mass loss is a  thorough understanding of the atmosphere of exoplanets. \cite{Vidal2003} detected an excess absorption of 15\% in Ly $\alpha$ by using transmission spectra of HD 209458b, which indicates that HD 209458b is surrounded by a highly extended hydrogen atmosphere. The following observations in Ly $\alpha$ further confirmed this conclusion \citep{Vidal2008, Ehrenreich2008}. Hydrodynamic models support the existence of the extended atmosphere of HD 209458b \citep{Yelle2004, Garcia2007, Murray2009, Guo2011, Guo2013, Guo2016}; however, the hot neutral hydrogen atoms (ENAs) produced by charge exchange between the stellar wind and planetary wind \citep{Holmstrom2008, Khodachenko2017} are requested because the highest velocity of the hydrogen atoms exceeds 100Km s$^-1$. Subsequent observations on HD 209458b discovered some heavy elements, such as C, O, Mg and Si \citep{Vidal2004, Vidal2013, Ballester2015, Linsky2010}. In addition, the magnetic field exists widely in the solar system and can have a great influence on the atmospheres of planets. The modeling of MHD shows that the magnetic fields modify the patterns of the atmosphere escape \citep{Trammell2014}, and a recent study on HD 209458b expresses that a magnetic field less than 1 Guass can trap the atmosphere in the dead zone around the equatorial surface \citep{Khodachenko2021_0}.


The observation in Ly$\alpha$ is a powerful method for detecting the atmosphere of exoplanets. However, the interstellar absorption obscures the signals of Ly$\alpha$ at a certain extent, and such observations must be operated by a space telescope. The He I 2$^3$S-2$^3$P triplet (namely, He 2$^3$S, 10830.33, 10830.25, and 10829.09 \AA) that avoids the shortage of Ly$\alpha$ observations \citep{Indriolo2009} is a new window for search the signals of a planetary atmosphere \citep{Seager2000, Oklopcic2018}. Excess absorption of He 2$^3$S has been confirmed in several close-in exoplanets \citep{Allart2018,Spake2018,Alonso2019,Czesla2022}. In order to interpret the He 2$^3$S absorption lines in these planets, \cite{Lampon2020, Lampon2021} used an isothermal atmosphere model. By varying the mass loss rate and temperature, they fitted the lines and found that the He/H ratio of HD 209458b, HD 189733b and GJ 3470b are 2/98, 0.8/99.2 and 1.5/98.5, respectively. 3D modeling of HD 209458b \citep{Khodachenko2021_0}, HD 189733b \citep{Rumenskikh2022}, and GJ 3470b \citep{Shaikhislamov2021} also support a low He/H ratio. However, a value close to that of Sun can fit the He 2$^3$S absorption lines of Wasp-107b \citep{Khodachenko2021_1, Wang2021_1} and Wasp-69b \citep{Wang2021_0}. Thus, the diverse properties of the H/He ratio remain to be further investigated.

As a well-studied exoplanet, the cause of the low helium abundance of HD 209458b needs to be explained because \cite{Lampon2020} only assume a low He/H ratio in order to fit the observation. The low value of He/H could be the consequence of a few possibilities. First, they could be related to planet formation. However, a full description of the process is beyond the scope of this paper. Another possibility is that the depletion of helium in the upper atmosphere is caused by the hydrogen-helium immiscibility layer in the planetary interior. For instance, the mechanism can decrease 15\% of helium in the upper atmosphere of Jupiter compared to the value of the protosolar nebula \citep{Wilson2010}. Nonetheless, the effect of the mechanism on Hot Jupiter is unclear. Here we propose a third possible explanation for the low He/H, namely, the helium is cannot escape as easily as hydrogen in a hydrodynamic escaping atmosphere due to the mass-dependent fractionation, which means that the insufficient flux of hydrogen can lead to a very low escaping fluxes for heavier species in the upper atmosphere \citep{Hunten1987}. For instance, for Earth-like planets, the escaping flux of oxygen in a water-dominated atmosphere is limited by that of hydrogen \citep{Guo2019}.

In this paper, we focus on how the fractionation of the mass affects the ratio of He/H in the upper atmosphere of HD 209458b by using a multi-fluid model. In fact, \cite{Guo2011} has shown the possibility of decoupling of H and H$^+$ in the atmosphere of HD 209458b. Therefore, because of the larger atomic mass of helium, the fractionation of mass in helium can be easier. In fact, \cite{Hu2015} explored the possibility of the formation of helium-dominated atmospheres due to  mass fractionation. However, they used a revised energy-limited equation to estimate the escaping fluxes of hydrogen and helium. Nevertheless, there is still a lack of quantitative investigations in the mass fractionation of helium.
Moreover, mass fractionation cannot be described via the single-fluid model as the velocities of all species are equal so the He/H ratio will remain constant in all altitudes in this regime. On the contrary, in a multi-fluid model, helium and hydrogen will run as separate fluids. In this situation, one can self-consistently study how much helium can escape from the atmosphere of the exoplanet via the drag of hydrogen instead of a presumed He/H ratio given at the lower boundary of the atmosphere. Such models are important not only for helium, but also for other heavy elements. This paper is organized as follows: Section \ref{sec:model_description} describes the equations and numerical methods. The structures of the atmosphere are described in Section \ref{subsec:atmosphere}. We discuss the fractionation of helium in Section \ref{subsec:fraction}. We fit the observation and obtained the best-fit parameters in Section \ref{subsec:observation}. In Section \ref{sec:Discussions}, we discuss our new numerical method and the effects of mass fractionation. We conclude with our results in Section \ref{sec:Conclusions}.

\section{Multi-fluid HD model} \label{sec:model_description}
We used the open-source MHD code PLUTO \citep{Mignone2007} to study the fractionation of He in a hydrogen-helium atmosphere. PLUTO aims to research the behavior of single-fluid. Here the atoms and ions of hydrogen and helium must be described as separate fluids. At the same time, the additional source terms, such as the heating, cooling, ionization, and chemical reactions of hydrogen and helium, must also be treated in the hydrodynamic equations. They are incorporated into PLUTO (see below). In addition, our main purpose is to determine how  fractionation affects the ratio of He/H in the upper atmosphere of HD 209458b. Thus, we only considered five kinds of species: H, H$^{+}$, He, He$^{+}$ and e$^-$. Molecular hydrogen and other heavy elements are neglected.

\subsection{The multi-fluid HD equations}\label{sec:HD_equations}

Our model includes the equations of number density, velocity, and pressure for H, He, H$^+$, He$^+$ and e$^-$. In an ionized atmosphere composed of atoms and ions, photoelectrons produced by stellar XUV irradiation distribute their energy via collision with thermal electrons. Ions obtain energy from thermal electrons via Coulomb collisions, while hydrogen atoms exchange energy with thermal electrons and ions by elastic and inelastic collisions. Thus, the heating by stellar XUV radiation is included in the equation of electron pressure. The equations of multi-fluid HD models are listed as follows:

\begin{equation}\label{eq:continue_1}
\frac{\partial n_ {t}}{\partial t}+\nabla\cdot(n_ {t}\textbf{u}_ {t})=S_ {t}-L_ {t},\ t=s,n
\end{equation}

\begin{equation}\label{eq:n_momt_2}
\frac{\partial \textbf{u}_ {n}}{\partial t}+ \textbf{u}_ {n}\cdot\nabla\textbf{u}_{n}
+\frac{\nabla p_ {n}}{m_{n}n_{n}}
=\sum_{t=all}\frac{C_{nt}n_{t}}{A_{n}}(\textbf{u}_{t}-\textbf{u}_{n})
+\frac{S_{n}}{n_{n}}(\textbf{u}_{s}-\textbf{u}_{n})
+a_{ext}
\end{equation}

\begin{equation}\label{eq:i_momt_3}
\frac{\partial \textbf{u}_ {s}}{\partial t}+\textbf{u}_ {s}\cdot\nabla\textbf{u}_{s}
+\frac{\nabla (p_{s}+p_{es})}{m_{s}n_{s}}
=\sum_{t=all}\frac{C_{st}n_{t}}{A_{s}}(\textbf{u}_{t}-\textbf{u}_{s})
+\frac{S_{s}}{n_{s}}(\textbf{u}_{n}-\textbf{u}_{s})
+a_{ext}
\end{equation}

\begin{equation}\label{eq:n_prs_4}
\begin{aligned}
\frac{\partial p_{n}}{\partial t}+
(\textbf{u}_{n}\cdot\nabla)p_{n}+\gamma p_{n}(\nabla\cdot&\textbf{u}_{n})                                                                                                                                                                                                                                                                                                                                                                                      
=\sum_{t=all}\frac{C_{nt}}{A_{n}+A_{t}}[2(p_{t}n_{n}-p_{n}n_{t})+
\frac{2}{3}m_{t}n_{n}n_{t}(\textbf{u}_{t}-\textbf{u}_{n})^2
]\\
&+\frac{S_{n}}{n_{s}}p_{s}-\frac{L_{n}}{n_{n}}p_{n}
+\frac{1}{3}S_{n}m_{n}(\textbf{u}_{s}-\textbf{u}_{n})^2\\
\end{aligned}
\end{equation}

\begin{equation}\label{eq:i_prs_5}
\begin{aligned}
\frac{\partial p_{s}}{\partial t}+
(\textbf{u}_{s}\cdot\nabla)p_{s}+\gamma p_{s}(\nabla\cdot&\textbf{u}_{s})                                                                                                                                                                                                                                                                                                                                                                                      
=\sum_{t=all}\frac{C_{st}}{A_{s}+A_{t}}[2(p_{t}n_{s}-p_{s}n_{t})+
\frac{2}{3}m_{t}n_{s}n_{t}(\textbf{u}_{t}-\textbf{u}_{s})^2
]\\
&+\frac{S_{s}}{n_{n}}p_{n}-\frac{L_{s}}{n_{s}}p_{s}
+\frac{1}{3}S_{s}m_{s}(\textbf{u}_{n}-\textbf{u}_{t})^2\\
\end{aligned}
\end{equation}

\begin{equation}\label{eq:e_prs_6}
\frac{\partial p_{e}}{\partial t}+
(\textbf{u}_{e}\cdot\nabla)p_{e}+\gamma p_{e}(\nabla\cdot\textbf{u}_{e})                                                                                                                                                                                                                                                                                                                                                                                      
=\sum_{t=all}\frac{C_{et}}{A_{e}+A_{t}}2(p_{t}n_{e}-p_{e}n_{t})
-\frac{L_{e}}{n_{e}}p_{e}+\frac{2}{3}(\mathscr{H}-\mathscr{L})
\end{equation}

\begin{table}[!hp]
\begin{threeparttable}[t]
\centering
\caption{Chemical reactions included in our model.}
\begin{tabular}{lll}
\toprule
  Reactions&Rates &References\\
\midrule
  $H+h\nu\to H^{+}+e^- $&  &\cite{Ricotti2002} \\
  $He+h\nu\to He^{+}+e^- $&  &\cite{Ricotti2002} \\
  $H+e^-\to H^{+}+e^-+e^-$ &$2.91\times 10^{-8}(\frac{1}{0.232+U})U^{0.39}exp(-U), U=13.6eV/E_{e}$  &\cite{Voronov1997}\\
  $He+e^-\to He^{+}+e^-+e^-$ &$1.75\times 10^{-8}(\frac{1}{0.180+U})U^{0.35}exp(-U), U=24.6eV/E_{e}$  &\cite{Voronov1997}\\
  $H^{+}+e^-\to H + h\nu$ &$4.0\times 10^{-12}(300/T_{e})^{0.64}$\tnote{(1)}  &\cite{Storey1995} \\
  $He^{+}+e^-\to He + h\nu$ &$4.6\times 10^{-12}(300/T_{e})^{0.64}$  &\cite{Storey1995} \\
  $H+He^{+}\to H^{+}+He$ &$1.25\times 10^{-15}{(300/T_{r})}^{-0.25}$\tnote{(2)}  & \cite{Glover2007}\\
  $H^{+}+He\to H+He^{+}$ &$1.75\times 10^{-11}{(300/T_{r})}^{0.75}exp(-128000/T)$  & \cite{Glover2007}\\
\bottomrule
\end{tabular}\label{tab:che_net}

\textbf{Notes.}
\begin{tablenotes}
\item[(1)]$T_e$ temperature of the electron
\item[(2)]$T_r = \frac{m_sT_t+m_tT_s}{m_s+m_t}$ is the reduced temperature where $s$ and $t$ represent the two reactants.
\end{tablenotes}
\end{threeparttable}
\end{table}

\begin{table}[!t]
\begin{center}
\begin{threeparttable}[t]
\centering
\caption{Collision rate coefficients  $C_{st}$ included in our model. }

\begin{tabular}{lll}
\toprule
  Species s,t &$C_{st}$&References\\

\midrule
  H, H$^+$ &  $2.67\times 10^{-10}T^{1/2}_{r}(1-0.083 log_{10}T_{r})^2\tnote{(1)}$&\cite{Schunk1980} \\
  He, He$^+$&  $3.50\times 10^{-10}T^{1/2}_{r}(1-0.093 log_{10}T_{r})^2$&\cite{Schunk1980}\\
  H, He&   $k_{B}T_{r}/(m_{amu}b)\tnote{(2)}\ $&\cite{Mason1970}\\
  H$^+$, He$^+$& $1.15\times T_{r}^{-3/2}$&\cite{Schunk1980}\\
  H$^+$, e$^-$&   $0.0299\times T_{e}^{-3/2}$&\cite{Schunk1980}\\
  He$^+$, e$^-$&  $0.0299\times T_{e}^{-3/2}$&\cite{Schunk1980}\\
\bottomrule

\end{tabular}\label{tab:cls_net}
\textbf{Notes.}\\
All of these $C_{st}$ are calculated by Equation (\ref{eq:cst_nu}).
\begin{tablenotes}
\item[(1)]$T_r = \frac{m_sT_t+m_tT_s}{m_s+m_t}$ is the reduced temperature.
\item[(2)]$k_B$ is Boltzmann constant, $m_{amu}$ is the mass in atomic mass units and $b$ is the binary diffusion coefficient.
\end{tablenotes}
\end{threeparttable}
\end{center}
\end{table}
where subscripts $n$, $s$, and $e$ denote neutral, ion, and electron fluid, respectively. In Equations (\ref{eq:continue_1})-(\ref{eq:e_prs_6}) $n$ is the number density, $\textbf{u}$ is the velocity, $p$ is pressure. $S_{t}$ and $L_{t}$ are the production and loss of particles. $m$ is the particle mass, and $C$ is the artificial collision rate coefficient that is expressed from the collision frequency. $A$ is the particle mass in atomic mass units ($m_{amu}$), $a_{ext}$ is the acceleration of fluid produced by external forces, and $\gamma$ is the adiabatic index. On the left side of the Equation (\ref{eq:i_momt_3}), the electric force $-\nabla p_{es}$ is treated as part of ion fluid pressure. Following \cite{Dong2017} $p_{es}$ can be expressed as

\begin{equation}\label{eq: QN ve}
p_{es} = \frac {q_{s}n_{s}}{en_{e}}p_{e}.
\end{equation}
which is the consequence of quasi-neutrality when neglecting the momentum of the electron.

Including the equation of the electron pressure resulted in a numerical difficulty in using the Riemann solver. We thus modified the process of the calculation by solving the equations of atoms, ions, and electrons separately, which will be discussed in Section \ref{subsec:num_method}.

Our model also includes the impact of the ionization electrons, charge exchanges between neutral and ions, recombination of ions and electrons , and other collisions. All chemical reactions included in this model are listed in Table \ref{tab:che_net}. The collision coefficients $C_ {st}$ for two species $s$ and $t$ we used in Equations (\ref{eq:continue_1})-(\ref{eq:e_prs_6}) are derived from collision frequency $\nu_{st}$ \citep{Schunk1980}. Following the momentum of conservation, the $C_{st}$ is defined as

\begin{equation}\label{eq:cst_org}
n_ {s}m_ {s} \nu_ {st}=n_ {t}m_ {t} \nu_ {ts}=C_ {st}m_ {amu}n_ {s} n_{t},
\end{equation}
where $m_{amu}$ is the atomic mass unit, $s$ and $t$ are the two collision species. Then, we can get $\nu_ {st}$ and $\nu_{ts}$ from $C_ {st}$

\begin{equation}\label{eq:cst_nu}
\begin{aligned}
\nu_ {st} = \frac{C_ {st}n_ {t}}{A_ {s}},
\nu_ {ts} = \frac{C_ {st}n_ {s}}{A_ {t}},
\end{aligned}
\end{equation}
It is convenient to include the collisional terms in momentum and energy equations. The collision coefficients are listed in Table \ref{tab:cls_net}. The ion-neutral collisions of the same element (.i.e, collisions between H and H$^+$ and He and He$^+$) are resonant collisions while the collisions between different elements (.i.e, collisions between H and He$^+$ and He and H$^+$) are nonresonant collisions.

In our model, the planetary and stellar tidal accelerations are included in the term of external forces. In this case $a_{ext}$ can be expressed as
\begin{equation}\label{eq:a_ext}
a_{ext}=-\frac{GM_{p}}{r^2}+\frac{3GM_{*}r}{a^3}
\end{equation}
where G is the gravitational constant, $M_{p}$ is the planet mass, $r$ is the altitude from center of the planet, $M_{*}$ is the stellar mass, $a$ is the semi-major axis. In Equation (\ref{eq:a_ext}) the first and second terms denote the acceleration produced by the planet and star, respectively.

In the process of photoionization, electrons obtain most of the energy of photoelectrons, which will be transferred to other species through the collision processes. Therefore, the heating of the atmosphere is only included in the electron pressure equation. Here we use an average heating efficiency $\eta$ to represent the different heating efficiency of photons with different frequencies. The heating term can be written as

\begin{equation}\label{eq:heat}
\mathscr{H}=\eta \sum_{\nu}(\sigma_{\nu}^{H}n_{H}+\sigma_{\nu}^{He}n_{He}) F_{\nu}e^{-\tau_{\nu}}
\end{equation}
and
\begin{equation}\label{eq:tau_get}
\tau_{\nu} = \int^{\infty}_{r}(\sigma_{\nu}^{H}n_{H}+\sigma_{\nu}^{He}n_{He})dr
\end{equation}

Where $\tau_{\nu}$ is optical depth, $\sigma_{\nu}^{H}$ and $\sigma_{\nu}^{He}$ are optical cross sections for H and He separately which are given by \cite{Ricotti2002} and $F_\nu$ is spectrum of irradiated stellar flux. We also included the $Ly\ \alpha$ cooling of hydrogen \citep{Murray2009}
\begin{equation}\label{eq:cooling}
\mathscr{L}=7.5\times 10^{-19}n_{e}n_{H}e^{-118348/T}ergcm^{-3}s^{-1}
\end{equation}

Because the mass of the electron is much smaller than other atoms and ions, we obtained the electron number density $n_e$ and velocity $u_e$ by the condition of quasi-neutrality \citep{Toth2012},

\begin{equation}\label{eq:QN_ne}
n_ {e} = \sum_{s=ions}\frac{q_ {s}n_ {s}}{e}
\end{equation}

\begin{equation}\label{eq:QN_ue}
u_{e}=u_{+} = \sum_{s=ions}\frac{q_ {s}n_ {s}u_ {s}}{en_ {e}}
\end{equation}
where $q_s$ is the electric charge of ion, $e$ is the electric charge of electron, $u_+$ is the mean velocity of all ion fluids and, $u_{e}$ is the velocity of the electron. Because the magnetic field is not included in this model, the effect of electric current is not considered.

\subsection{Model structure and Numerical method}\label{subsec:num_method}
Equations (\ref{eq:continue_1})-(\ref{eq:e_prs_6}) are solved using PLUTO \citep{Mignone2007}, and a third-order total variation diminishing (TVD) Runge-Kutta schemes method is implemented into PLUTO for the time integration. The integration of time from $t^{n}$ to $t^{n+1}$ is expressed as

\begin{equation}\label{eq:RK3_1}
\textbf{U}^{*}=\textbf{U}^{n}+\Delta t\textbf{H}(\textbf{U}^{n}),
\end{equation}

\begin{equation}\label{eq:RK3_2}
\textbf{U}^{**}=\frac{1}{4}(3\textbf{U}^{n}+\textbf{U}^{*}+\Delta t\textbf{H}(\textbf{U}^{*})),\\
\end{equation}

\begin{equation}\label{eq:RK3_3}
\textbf{U}^{n+1}=\frac{1}{3}(\textbf{U}^{n}+2\textbf{U}^{*}+2\Delta t\textbf{H}(\textbf{U}^{**})),
\end{equation}
where \textbf{U} is $n, \textbf{u}$ and $p$ for H, He, H$^+$, He$^+$ and e$^-$. The residual term $\textbf{H}$ is the sum of flux terms (or advection terms) $\textbf{L}_{F}$ and source terms $\textbf{L}_{C}$ namely,
\begin{equation}\label{eq:H_C}
\textbf{H}=\textbf{L}_{F}+\textbf{L}_{C}.
\end{equation}
Note that all right-hand side terms in Equations \ref{eq:continue_1}-\ref{eq:e_prs_6} are included in $\textbf{L}_{C}$. We applied the semi-implicit method of \cite{Garcia2007} to calculate $\textbf{H}$,

\begin{equation}\label{eq: semi-implicit}
[I-\frac{\delta \textbf{L}_{C}}{\delta \textbf{U}}]\textbf{H}(\textbf{U})=
  \textbf{L}_{C}(\textbf{U})+ \textbf{L}_{F}(\textbf{U}),
\end{equation}
in which the Jacobian matrix $\frac{\delta \textbf{L}_{C}}{\delta \textbf{U}}$ is treated numerically \citep{Toth2012}.

There are three kinds of fluids in our model, i.e., neutral fluids for H and He, ion fluids for H$^{+}$ and He$^{+}$ and e$^-$ fluid.  For the neutral fluids, the Riemann solver of PLUTO can solve them directly as in \cite{Toth2012}. For the ion fluids, the additional electric field force $-\nabla p_{es}$ impedes the use of the Riemann solver.
The flux terms of the ion fluids can be expressed as
\begin{equation}\label{eq:F_s}
\textbf{L}_{Fs} =
\begin{pmatrix}
-\nabla\cdot(n_{s}\textbf{u}_{s})\\
-(\textbf{u}_ {s}\cdot\nabla\textbf{u}_{s}+\frac{\nabla (p_{s}+p_{es})}{m_{s}n_{s}})\\
-((\textbf{u}_{s}\cdot\nabla)p_{s}+\gamma p_{s}(\nabla\cdot\textbf{u}_{s}))                                                                                                                                                                                                                                                                                                                                                                                      
\end{pmatrix}
\end{equation}
The terms in Equation ({\ref{eq:F_s}}) represent number density $n_s$, velocity $u_s$ and pressure $p_s$ flux terms of the ion fluids, respectively. In fact, the flux terms in the form
\begin{equation}\label{eq:Euler}
\textbf{L}_{F}=
\begin{pmatrix}
-\nabla\cdot(n\textbf{u})\\
-(\textbf{u}\cdot\nabla\textbf{u}+\frac{\nabla (p)}{mn})\\
-[(\textbf{u}\cdot\nabla)p+\gamma p(\nabla\cdot\textbf{u})]                                                                                                                                                                                                                                                                                                                                                                                      
\end{pmatrix}
\end{equation}
can be solved directly by the Riemann solver. By comparing the Equations (\ref{eq:Euler}) and (\ref{eq:F_s}), we cannot solve Equation (\ref{eq:F_s}) with the Riemann solver due to the additional term $p_{es}$. However, the Riemann solver can be applied if $p$ in Equation (\ref{eq:Euler}) is replaced by $p_{s}+p_{es}$. Under this condition, the flux terms are expressed as

\begin{equation}\label{eq:F_s_nu}
\textbf{L}_{Fs(\rho,u)}=
\begin{pmatrix}
-\nabla\cdot(n_{s}\textbf{u}_{s})\\
-(\textbf{u}_ {s}\cdot\nabla\textbf{u}_{s}+\frac{\nabla (p_{s}+p_{es})}{m_{s}n_{s}})\\
-((\textbf{u}_{s}\cdot\nabla)(p_{s}+p_{es})+\gamma (p_{s}+p_{es})(\nabla\cdot\textbf{u}_{s}))                                                                                                                                                                                                                                                                                                                                                                                      
\end{pmatrix}
\end{equation}
Equation (\ref{eq:F_s_nu}) can be solved by the Riemann solver and we can obtain the flux terms of $n_s$, $u_s$ and $p_s+p_{es}$. The integration of time in the Riemann solver is explicit, thus the flux terms of $n_s$ and $u_s$ in the time t$^{n+1}$ only depend on the values at t$^{n}$. Thus, in the process of solving Equation (\ref{eq:F_s_nu}) the correct values of $n_s$ and $u_s$ are obtained. At the same time, we used an incorrect flux term for $p_s$ in the pressure equation so that the value of $p_s$ obtained via the flux term of $p_s+p_{es}$ is discarded. In order to calculate $p_s$, we introduce a new density $n'$ to replace the $n$ in Equation (\ref{eq:Euler}). The flux terms are modified again as
\begin{equation}\label{eq:F_s_p}
\textbf{L}_{Fs(p)}=
\begin{pmatrix}
-\nabla\cdot(n'\textbf{u}_{s})\\
-(\textbf{u}_{s}\cdot\nabla\textbf{u}_{s}+\frac{\nabla p_{s}}{m_{s}n'})\\
-[(\textbf{u}_{s}\cdot\nabla)p_{s}+\gamma p_{s}(\nabla\cdot\textbf{u}_{s})]                                                                                                                                                                                                                                                                                                                                                                                      
\end{pmatrix}
\end{equation}
which are solved by the Riemann solver again. As expressed above, due to the explicit characteristic in the time integration we can obtain the $p_{s}$ in the process of solving Equation (\ref{eq:F_s_p}), and an artificial value of $n'$ in Equation (\ref{eq:F_s_p}) is permitted in mathematics because $p_{s}$ is independent of $n'$ when the Equation (\ref{eq:F_s_p}) is solved.

Nevertheless, a reasonable value for $n'$ is necessary. The escape of ions is driven by both the pressure of ions and electrons. Thus they can be divided into two parts of which one is driven by the pressure of ions and the corresponding number density is denoted by $n'$. Another is driven by the electron pressure with number density $n_{es}$. Finally, the number density is the sum of the two parts, namely, $n_s =n'+n_{es}$. In fact, the escaping fluxes driven by $p_s$ and $p_{es}$ are proportional to the ratios of $p_{s}/p_{es}$, namely, $n'/n_{es}=p_{s}/p_{es}$. Then the number density $n'$ of the fluid driven by pressure $p_s$ is given by


\begin{equation}\label{eq:c_match}
n' = n_ {s}\frac{p_{s}}{p_ {s}+p_{es}}.
\end{equation}

Only the pressure of the electron remains to be calculated due to the assumption of quasi-neutrality. As manipulated above, we constructed the flux terms of $\rho_{e}'$ and $\textbf{u}_e$ where $\textbf{u}_{e}$ is obtained by Equation (\ref{eq:QN_ue}). For the density of electrons, an arbitrary value is acceptable. Here we applied a similar method to define the $\rho_{e}'$,

\begin{equation}\label{eq:c_e_mt}
\rho_{e}' = \sum_{s=ions}(n_{s}m_{s})\frac{p_{es}}{p_ {s}+p_{es}}
\end{equation}
which is equal to the summation of all mass density driven by the electric field force $-\nabla p_{es}$. Finally, the electron pressure is obtained by

\begin{equation}\label{eq:F_e_p}
\textbf{L}_{Fe(p)}=
\begin{pmatrix}
-\nabla\cdot(\rho_e'\textbf{u}_{e})\\
-(\textbf{u}_{e}\cdot\nabla\textbf{u}_{e}+\frac{\nabla p_{e}}{\rho_e'})\\
-[(\textbf{u}_{e}\cdot\nabla)p_{e}+\gamma p_{e}(\nabla\cdot\textbf{u}_{e})].                                                                                                                                                                                                                                                                                                                                                                                      
\end{pmatrix}
\end{equation}

The verification of this Riemann split method is shown in Appendix \ref{app:code}. The Riemann solver that we used in our work is HLL which is implemented in PLUTO. In fact, one can use different Riemann solvers for different requirements.

\subsection{Isothermal multi-fluid test model for the atmosphere of the Earth-like planet}\label{subsec:iso_test}

\begin{figure}[ht!]
\plotone{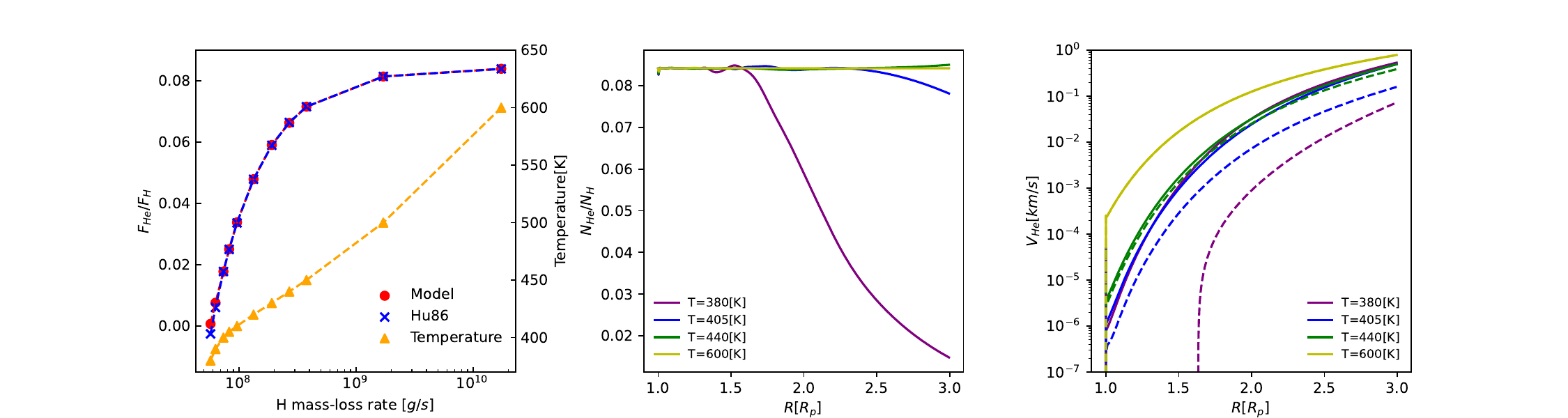}
\caption{Isothermal multi-fluid test result for the atmosphere of Earth-like planet. The left panel shows how $F_{He}/F_{H}$ evolves with mass-loss rate of hydrogen and their corresponding temperature (orange). Red and blue points show the ratio of escape flux of helium to that of hydrogen given by our model and Hunten (1987), respectively. The middle panel shows ratio of number density of helium to that of hydrogen for four cases of temperature = 380K, 405K, 440K and 600K. The right panel shows the velocities for hydrogen (solid lines) and helium (dashed lines) for the same four cases as middle panel. The velocities of helium and hydrogen are almost the same when the temperature is 600K due to sufficient collision between hydrogen and helium. For the case of T=380K, there is some slight negative velocity (absolute value less than 1cm/s) when R=1.0 to R=1.6.}
\label{fig:iso_aly}
\end{figure}

In order to test the collision of H and He, we ran a multi-fluid isothermal (setting polytropic exponent $\gamma = 1.0001$) model and compared the result with the analytical result given by \cite{Hunten1987}. Considering that this analytical result is suitable for an earth-like planet, we simulated the escape of a planet with the mass and radius. For simplicity, we run a two-fluid model constituted of neutral fluids of hydrogen and helium. The mass-loss rates of hydrogen fluid are modified by varying the temperature of the atmosphere. Following \cite{Hunten1987} the heavy species can be dragged by lighter particles through collisions if the mass of the heavy species is smaller than the crossover mass. The crossover mass and the escaping flux of helium are
\begin{equation}\label{eq:mc}
m_{c}=m_{H}+\frac{k_{B}TF_{H}}{bg_{0}X_{H}}
\end{equation}
and
\begin{equation}\label{eq:Flux_2}
F_{He}=\frac{X_{He}}{X_{H}}F_{H}(\frac{m_{c}-m_{He}}{m_{c}-m_{H}})
\end{equation}
where $m_{c}$ is crossover mass, $m_{H}$ and $m_{He}$ are the mass of H and He, $k_{B}$ is Boltzman constant, T is temperature, $F_{H}$ and $F_{He}$ are the escaping number density flux of H and He and $b$ is binary diffusion coefficient \citep{Mason1970}. Meanwhile,
$X_H=n^{0}_{H}/(n^{0}_{H}+n^{0}_{He})$ and $X_{He}=n^{0}_{He}/(n^{0}_{H}+n^{0}_{He})$ are the mixing ratio of H and He at the lower boundary.
The escaping flux of He and crossover mass can be obtained after the escaping flux of H is given. It is clear from (\ref{eq:Flux_2}) that the He can be dragged by H only if $m_c>m_{He}$ or vice versa and the ratios of $F_{He}/F_{H}$ are functions of $m_c $. At the same time, we used our model to calculate the escaping fluxes of H and He, namely:
\begin{equation}
F_{H,He} = \frac{\dot{M}_{H,He}}{m_{H, He}R_p^2}
\label{eq:f_H}
\end{equation}
where $\dot{M}_{H, He}$ are the mass-loss rate of hydrogen and helium, $m_{(H,He)}$ are the masses of H and He, $R_p$ is the radius of the planet. Meanwhile, the $F_{He}/F_{H}$ of our model is given by

\begin{equation}\label{eq:F0_F1}
\frac{F_{He}}{F_{H}}=\frac{\dot{M}_{He}/m_{He}}{\dot{M}_{H}/m_{H}}.
\end{equation}

As shown in the left panel of Figure \ref{fig:iso_aly}, the ratio of escape flux of helium to that of hydrogen in our model is consistent with that predicted by \cite{Hunten1987} quite well, which indicates the correctness of our model. The mass-loss rates increase roughly exonentially with the increase in temperature. The mixing ratio of helium can keep constant with the increase of the temperature as shown by the middle panel of Figure \ref{fig:iso_aly}. Meanwhile, the velocities of helium are dragged higher by the fluid of hydrogen with the increasing hydrogen mass-loss rate as shown in the right panel of Figure \ref{fig:iso_aly}. When the temperature drops to 380 K, the mixing ratio of helium drops dramatically. In this case, the escaping helium decreases to a fairly low level simultaneously. There are even some slight negative velocities at lower altitudes, probably due to the limitations of the code accuracy. In this extreme case, the $\frac{F_{He}}{F_{H}}$ predicted by \cite{Hunten1987}(blue dashed line in left panel of Figure \ref{fig:iso_aly}) is zero, which is consistent with that of our model ($7\times 10^{-4}$).

\section{Results} \label{sec:results}

\subsection{model settings} \label{subsec:setting}
We choose HD 209458b as a benchmark to explore the fractionation of He. The parameters of HD 209458 system are set as follows: the mass of the planet is  $0.69M_J$, the radius $1.36R_J$, the mass of the star $ 1.119\ M_{\odot}$, the semi-major axes $a = 0.04707$AU. They are the same as that of \cite{Lampon2020}.

Our 1D multi-fluid hydrodynamic model has 1024 grids from 1-10 $R_{p}$. The stretched ratio $dr_{n}/dr_{n-1}$ is 1.0062 where $dr_{n}$ is the length of the $n$th grid. Thus, most grid points are accumulated in the lower regions of the atmosphere. The values at the outer boundary are extrapolated linearly. The values at the lower boundary are different for different species. For the neutral fluid, the number densities of H and He are fixed as $1.0\times 10^{13}\ cm^{-3}$ and $8.46 \times 10^{11}\ cm^{-3}$. The ratio of $n_H/n_{He}$ is 0.0846, which is almost consistent with that of the Sun \citep{Asplund2009}. The number densities of ion fluid at the lower boundary are imposed to equal the value of the first grid point. We assumed that the temperatures at the lower boundary are the same for all fluids due to sufficient collisions, which is 1500 K and approximately equivalent to the effective temperature of HD 209458b. In such conditions, the pressure at the lower boundary is about 2 $\mu bar$, which is close to the homopause pressure given by \cite{Garcia2007}. The velocities of all fluids are given by equivalent extrapolation.

The spectral type of HD 209458 is similar to that of the Sun; therefore, we set a fiducial XUV flux (hereafter $F_{0}$) which is the same as that of the quiet Sun. The corresponding spectrum is obtained from \cite{Guo2016}($F_{0} = 2.62\ erg\ cm^2\ s$ at 1 AU in the wavelength range of 1\AA-912\AA), which is divided into 53 bins (1-912 \AA). In fact, the activity of the star can result in a variation in the XUV flux. In addition, the heating efficient $\eta$ in our model is a free parameter. Therefore, we calculated many cases by varying the XUV flux and heating efficiency. These models are composed of three groups. In group A, $\eta$ is fixed at 0.2 which is close to that given by \cite{Shematovich2014} and $F_{XUV}$ ranges from 0.60-4.0 times fiducial value. In group B, $F_{XUV}$ is fixed at 1.80 $F_{0}$ and $\eta$ ranges from 0.10-0.5. Group C is similar to group B except the XUV flux is the fiducial value (Table \ref{tab:model_para}).

\begin{table}[!t]
\begin{center}
\begin{threeparttable}[t]
\caption{Parameters and the model results for groups A-C. The crossover masses $m_c$ are calculated from Equation (\ref{eq:mc}) in unit of relative atom mass unit ($A_r$) when temperature set to balance temperature 1500K.} The $F_{He}/F_H$ are escaping flux ratio of He to H.

\begin{tabular}{ccccccc}
\toprule
 N& Group& $F_{XUV}$& Heating efficiency $\eta$& Mass-loss rate $\dot{M}$& $m_c(T=1500K)$& $F_{He}/F_{H}$\\
 & & [$F_{0}$]& & $[10^{10}g/s]$& $[A_{r}]$\\

\midrule
  1&A& 0.60& 0.20& 0.595& 3.01& $1.11\times 10^{-4}$\\
  2&A& 0.75& 0.20& 0.785& 3.63& $2.09\times 10^{-3}$\\
  3&A& 1.00& 0.20& 1.09& 4.52& $1.18\times 10^{-2}$\\
  4&A& 1.50& 0.20& 1.67& 6.02& $3.05\times 10^{-2}$\\
  5&A& 1.80& 0.20& 2.01& 6.87& $3.89\times 10^{-2}$\\
  6&A& 2.00& 0.20& 2.24& 7.44& $4.32\times 10^{-2}$\\
  7&A& 2.50& 0.20& 2.83& 8.92& $5.07\times 10^{-2}$\\
  8&A& 3.00& 0.20& 3.43& 10.5& $5.48\times 10^{-2}$\\
  9&A& 4.00& 0.20& 4.64& 13.7& $5.91\times 10^{-2}$\\
\hline
  10&B& 1.80& 0.10& 0.959& 4.15& $7.27\times 10^{-3}$\\
  11&B& 1.80& 0.15& 1.49& 5.58& $2.54\times 10^{-2}$\\
  12&B& 1.80& 0.20& 2.01& 6.87& $3.89\times 10^{-2}$\\
  13&B& 1.80& 0.30& 3.06& 9.53& $5.27\times 10^{-2}$\\
  14&B& 1.80& 0.40& 4.12& 12.3& $5.79\times 10^{-2}$\\
  15&B& 1.80& 0.50& 5.13& 14.9& $6.06\times 10^{-2}$\\
\hline
  16&C& 1.00& 0.14& 0.716& 3.41& $9.34\times 10^{-4}$\\
  17&C& 1.00& 0.17& 0.909& 4.00& $5.42\times 10^{-3}$\\
  18&C& 1.00& 0.20& 1.09& 4.52& $1.20\times 10^{-2}$\\
  19&C& 1.00& 0.30& 1.68& 6.04& $3.05\times 10^{-2}$\\
  20&C& 1.00& 0.40& 2.25& 7.47& $4.31\times 10^{-2}$\\
  21&C& 1.00& 0.50& 2.82& 8.91& $5.07\times 10^{-2}$\\
\bottomrule

\end{tabular}\label{tab:model_para}
\end{threeparttable}
\end{center}
\end{table}

\subsection{The Structure of the atmosphere}\label{subsec:atmosphere}

\begin{figure}[ht!]
\plotone{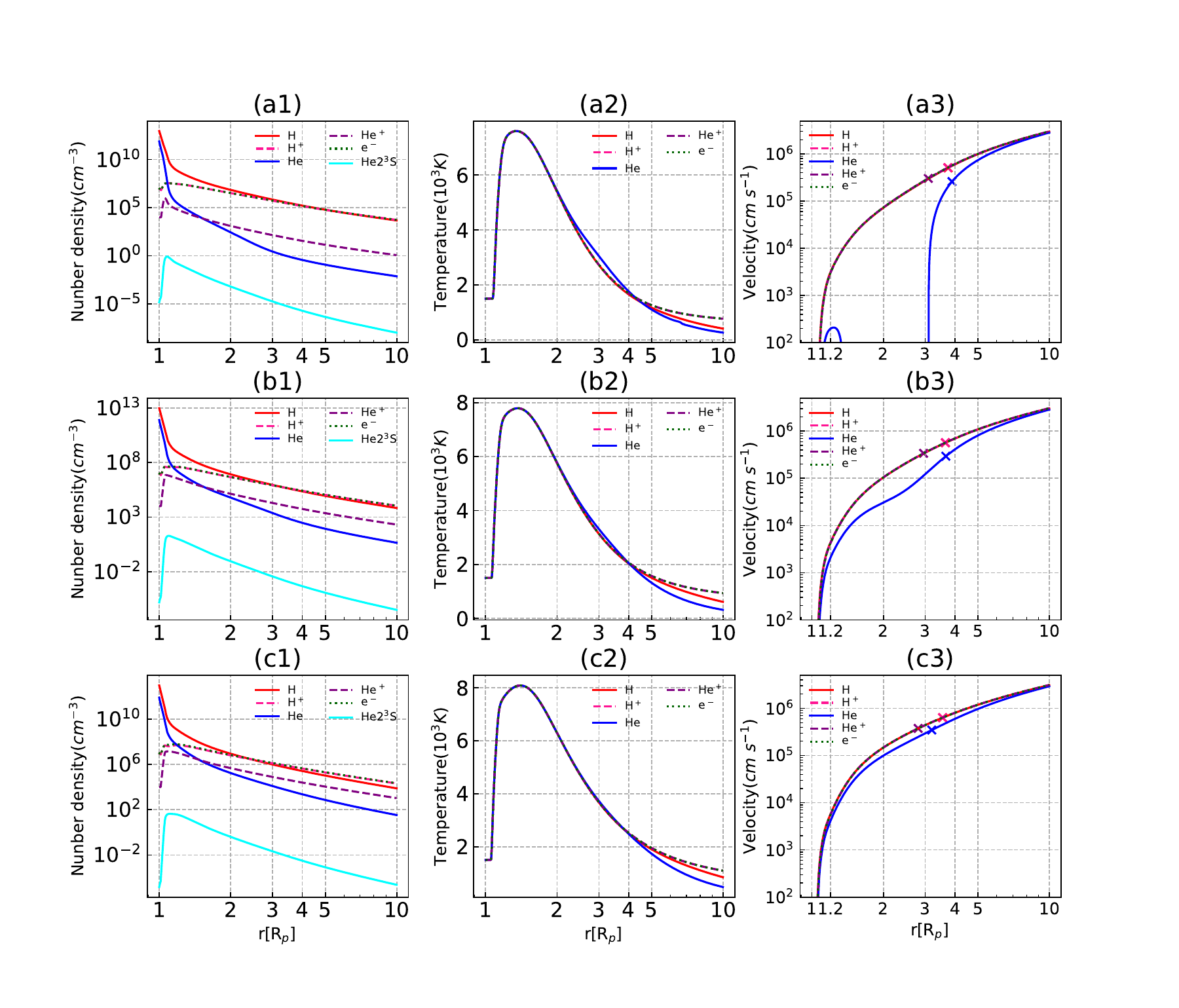}
\caption{The atmosphere structure of HD 209458b. The Panels (a1)-(a3) show the number density, temperature and velocity for H, He, H$^+$, He$^+$ and electron for the case with $\eta$ = 0.20 and $F_{XUV}$ =  0.60 $F_{0}$. The number density of He2$^3$S is also shown. The minus part of panel (a3) can be seen in Figure \ref{fig:cvg_ncp}. All species of this case are transonic and the corresponding sonic points are shown by "x" marks with the same color as their fluids. Note that the sonic points of H and H$^+$ are overlapped due to the high coupling between them. The middle and lower panels are the same as panel (a), but the XUV fluxes are 1.0 and 1.8 times $F_{0}$ respectively, and all species of them are transonic too. Meanwhile, the Roche sphere is about 4.2 $R_p$ in our model.} \label{fig:struc}
\end{figure}

\begin{figure}[ht!]
\plotone{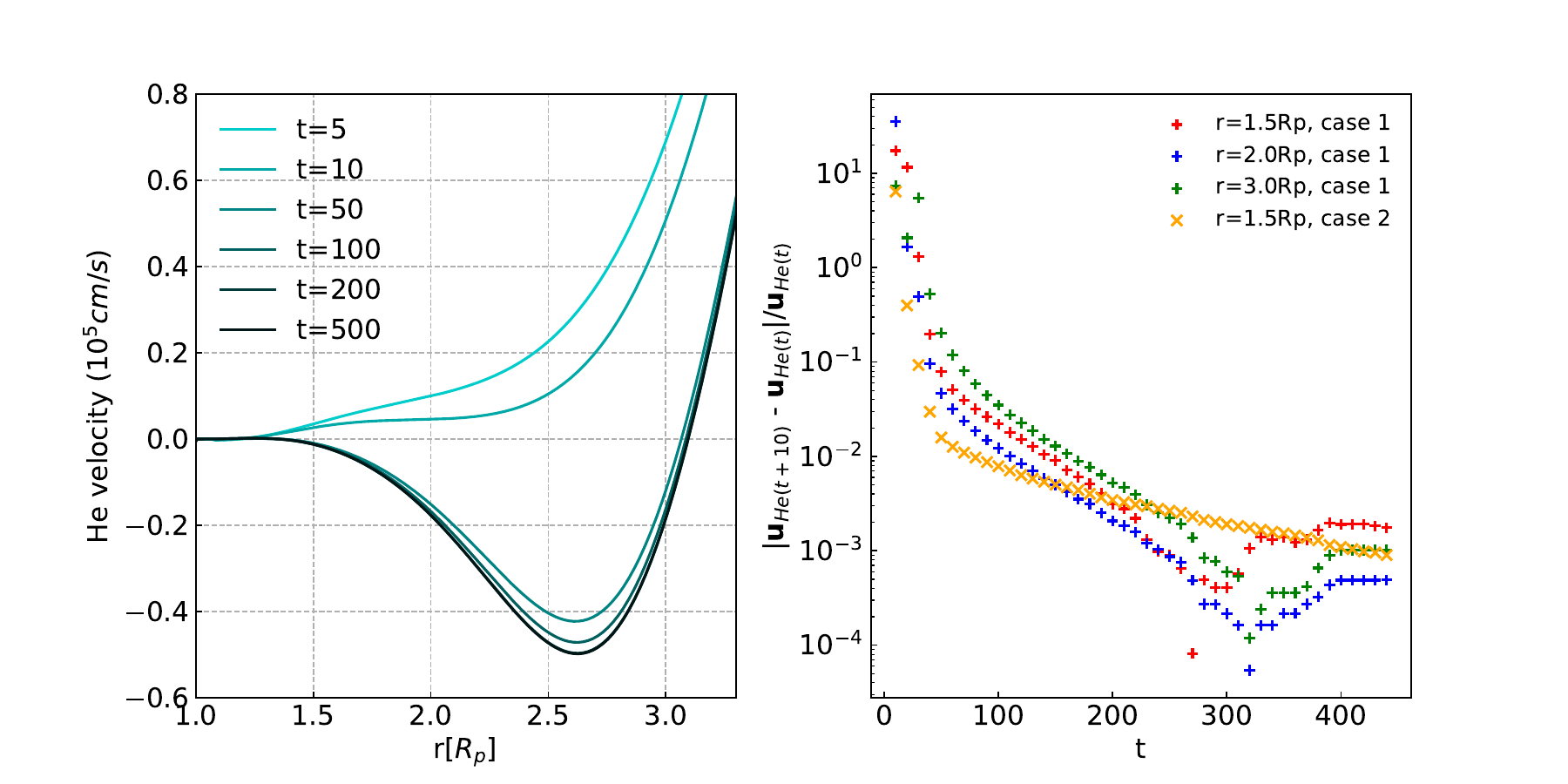}
\caption{Convergence of case 1 in group A of Table \ref{tab:model_para}. The left panel shows the velocity of helium in dimensionless integral times whose unit is about $10^5$ s. Here the velocities of t=500 are almost overlapped with the velocities of t=200. The right panel shows the variation of $|d\textbf{u}|/\textbf{u}$ of every 10 units of time (the corresponding integral steps is about $10^5$) at several altitudes of cases 1 and 2.}
\label{fig:cvg_ncp}
\end{figure}

Figure \ref{fig:struc} shows the structures of the atmosphere for some cases of group A. The XUV fluxes are 0.60, 1.0, and 1.8 times F$_{0}$. The profiles of the number density are shown in the panels (a1)-(c1). We can see that the number density of He decreases faster than that of H with the increase of the altitude, which is due to the fact that the spectra shorter than 504$\AA$ occupied most of the XUV flux so that the atomic helium encounters a stronger ionization than H. It is clear from the panels (a2),(b2) and (c2) that the temperature first rises to about 8000 K and decreases dramatically. The decrease in temperature is caused mainly by the expansion of the atmosphere. The temperatures of all species have similar profiles until about 3 R$_{p}$. Beyond 3 R$_{p}$ the decoupling is evident for atomic helium. The velocities of all species couple well for the case of 4 times F$_{0}$. A velocity decoupling between atomic helium and other species happens with the decrease of XUV flux. For the case of F$_{XUV}$=0.60F$_{0}$ the behavior of atomic helium is prominently different from other species. The velocity of atomic helium is even negative in the regions of $1.2R_{p}<r<3R_{p}$. In higher altitudes, the decoupling disappears due to the tidal forces of the host star. This negative velocity is not expected. We show the evolution of velocity with the integral time in Figure \ref{fig:cvg_ncp}. It should be noticed that the velocities are almost unchanged after 200 integral time units, and the $|d\textbf{u}|$ converges to one-thousandth of the velocities after 400 integral time units. Thus, the calculation results of both velocity and $|d\textbf{u}|/\textbf{u}$ converged to the stable status. A possible reason for the negative velocity is that the neutral helium fluid is in the quasi-hydrostatic status rather than hydrodynamic status, and such calculations are not suitable for PLUTO. This case needs to be explored further.

\begin{figure}[ht!]
\plotone{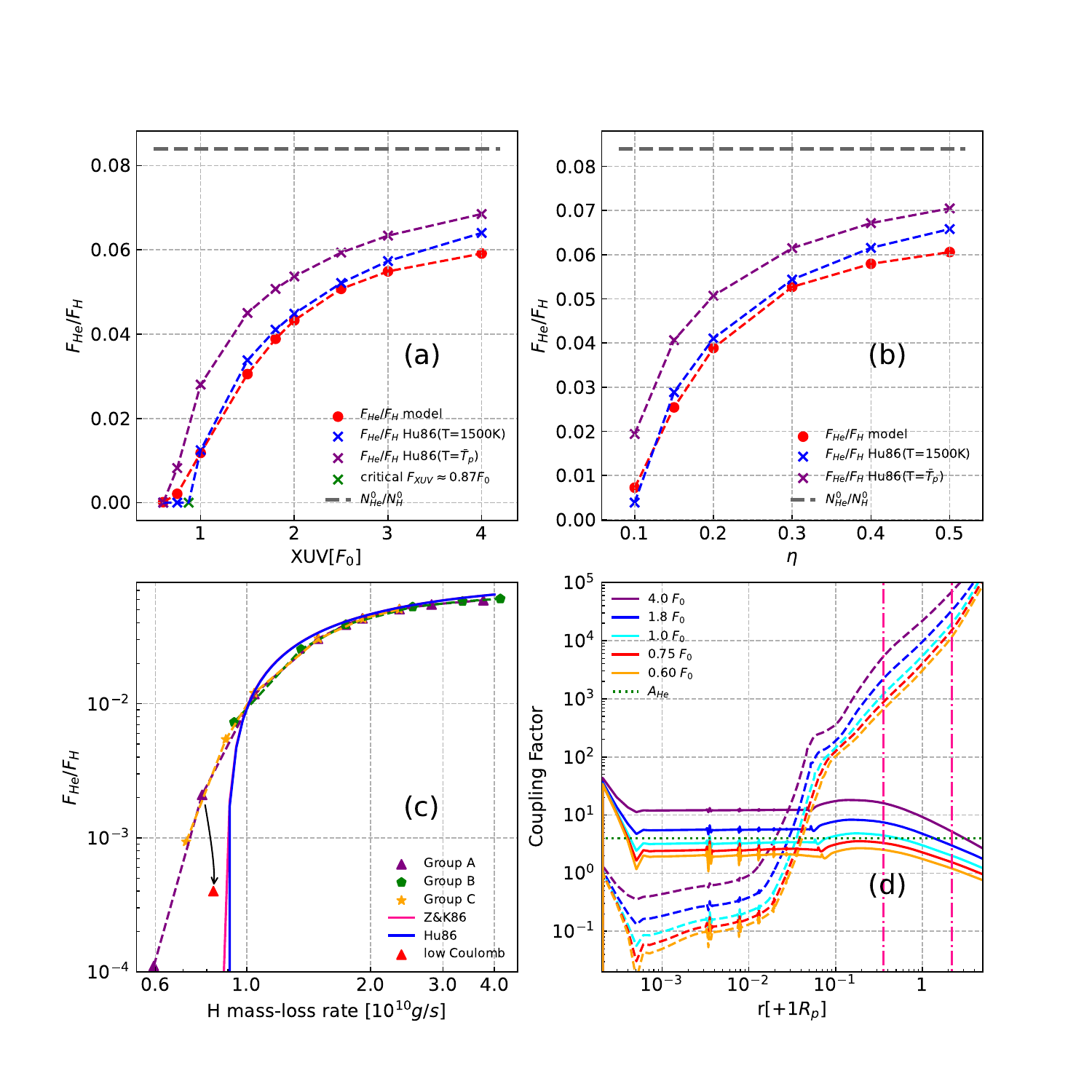}
\caption{The mass fractionation evolves with $\eta$ and $F_{XUV}$. Panels (a) and (b) show the ratio of $F_{He}/F_{H}$ of group A and B, respectively. Blue and purple points represent the $F_{He}/F_{H}$ given by Equation (\ref{eq:Flux_2}) when T=1500 K and T=$\bar{T}_p$, respectively. Red points represent the values of $F_{He}/F_{H}$ of our models. Note that $F_{He}$ and $F_{H}$ are the sum of neutral atoms and ions. As a comparison, the ratio of number density of He and H in the lower boundary are shown as a grey dash line.  Panel (c) shows the $F_{He}/F_{H}$ as a function with the mass loss rate of hydrogen $\dot{M}_{H}$. Purple, green, and orange points represent the models of groups A-C. The trends of the three groups are almost consistent. The pink and blue lines show the isothermal analytical results of \cite{Zahnle1986} and \cite{Hunten1987} when T=1500 K, respectively. Note that they are indistinguishable in the coupled regimes. The red triangle represents the case of $F_{XUV} = 0.75F_{0}, and \ \eta = 0.20$, but the Coulomb collision is decreased to $10^{-3}$ of its original value. In the panel (d), the solid and dash lines denote the coupling factor $\mathscr{C}_{st}$ of He-H and He$^+$-H$^+$  for some cases in group A of Table \ref{tab:model_para}, respectively. The dotted green line shows the relative atomic mass of He. For convenience to compare with Figure \ref{fig:he_decouple}, the area between these two pink doted-dashed lines represents the minus He velocity region of case 1 in Table \ref{tab:model_para}.
}\label{fig:mf_aly}.
\end{figure}

\begin{figure}[ht!]
\plotone{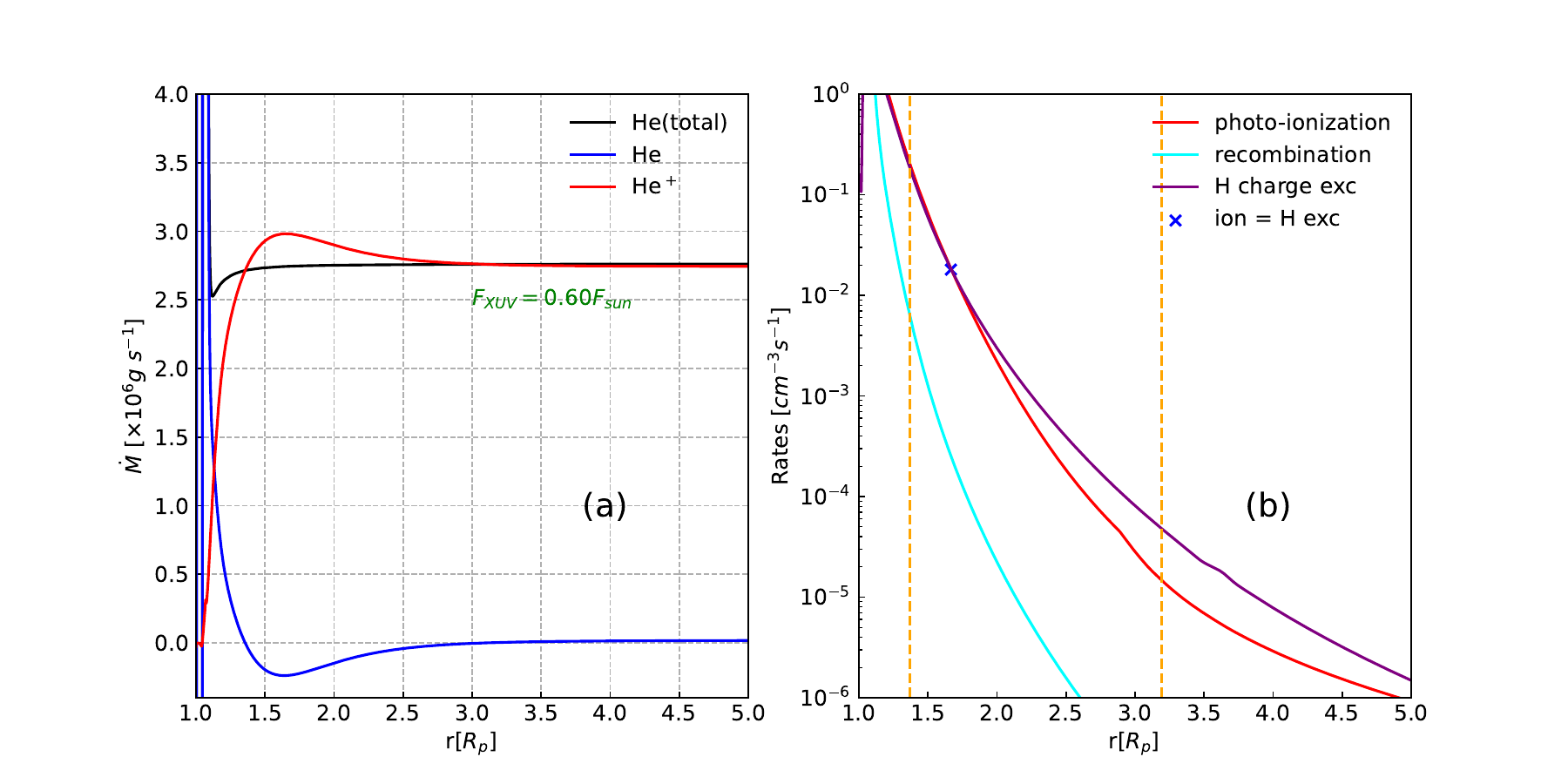}
\caption{Analysis of the decouple case when  $\eta = 0.20$ and $F_{XUV}=0.60$. Panel (a) shows the mass-loss rates of helium. Panel (b) shows the mass transfer between He and He$^+$ via chemical reactions. The red and purple lines denote the loss and the production of He from photoionizatio and the charge exchange between H and He$^+$. The cyan line denotes the production of He from recombination of He$^+$ and e$^-$. Blue x is the location where the photoionization rates equal to charge exchange rates between H and He$^+$. The production of He$^+$ from electron collision ionization and charge exchange between He and H$^+$ is too little compared with that of photoionization, so we do not put these in this panel. The negative velocities occur in the region between two orange lines.}
\label{fig:he_decouple}
\end{figure}

\subsection{Mass fractionation between He and H}\label{subsec:fraction}
The escape of helium plays a key role in simulating the absorption of helium triplets. As presented by \cite{Lampon2020}, a low mixing ratio of helium (about 2\%, which is fairly lower than the value of the Sun) is required in order to explain the absorption of helium triplet. Our simulations show that the mixing ratios of all models are lower than the solar value (0.086) (Table  \ref{tab:model_para}). Our results also indicate that a mixing ratio of 0.039 for He/H in the upper atmosphere of HD 209458b is the best-fit model for the absorption of helium triplets (Section \ref{subsec:observation}). A difference from the results of \cite{Lampon2020} is that the low He/H ratio in the upper atmosphere is not set artificially but as a result of the mass fractionation of helium.

We compared the $F_{He}/F_{H}$ of our model with those of \cite{Hunten1987}. The escaping flux of $F_{He}$ predicted by \cite{Hunten1987} can be obtained via Equations (\ref{eq:mc}) and (\ref{eq:Flux_2}). It is convenient to apply the equations of \cite{Hunten1987}. However, the temperature in a non-isothermal atmosphere needs to be defined firstly. In fact, the equations of \cite{Hunten1987} describe the friction of particles due to binary diffusion. Thus, here we used a pressure averaged temperature \citep{Koskinen2013}, which is calculated from the lower boundary to the altitude where the number density of atomic helium is equal to that of helium ions. The average temperature is defined as
\begin{equation}\label{eq:ave_tmp}
\bar{T}_{p}=\frac{\int^{p_{2}}_{p_{1}}T(p)d(lnp)}{ln(p_{2}/p_{1})}
\end{equation}
where $T(p) = \frac{m_HT_H+m_{He}T_{He}}{m_H+m_{He}}$ is reduced temperature, $p_1$ is the pressure at the lower boundary, and $p_2$ is the pressure at the location of n(He)=n(He$^{+}$). In fact, the altitude of n(He)=n(He$^{+}$) denotes the location above which the dominant particles in the atmosphere are ions. At the same time, we also calculated the cases that the temperature is equal to that of the lower boundary.

The crossover mass increases with the increase of the XUV flux and heating efficiency (Colume 6 of Table \ref{tab:model_para}). As a consequence, the fractionation of helium appears an opposite trend (panel (a) and (b) in Figure \ref{fig:mf_aly}). One can see that the escaping fluxes of helium are dependent on temperature. The $F_{He}/F_H$ predicted by Equation (\ref{eq:Flux_2}) is higher than ours if T=$\overline{T_{p}}$ ($\overline{T_{p}}\sim$ 3000-4000 K, purple points in panels (a) and (b) of Figure \ref{fig:mf_aly}) while those of T=1500 K are more consistent with ours (blue points in panels (a) and (b) in Figure \ref{fig:mf_aly}). In fact, the collision rates between H and He increase with the increase in temperature as shown in Table \ref{tab:cls_net}. Thus, a higher temperature causes a larger crossover mass, and thus higher escaping flux for helium. The critical value of $F_{XUV}$ that is defined in the condition of $m_c=m_{He}$ is 0.87$F_{0}$ in group A. The equations of \cite{Hunten1987} indicate that the $F_{He}$ will decrease to zero if $m_c<m_{He}$. In fact, the condition of $m_c<m_{He}$ is controlled mainly by the escaping flux of hydrogen. Thus, one should notice that an uncoupled regime will occur if the mass-loss rates of hydrogen (or XUV radiation) is lower than the critical value. We also found that the $F_{He}/F_{H}$ ratios ultimately depend on the mass loss rates of hydrogen. The panel (c) of Figure \ref{fig:mf_aly} shows that the effect of the fractionation of helium increases with the decrease of $\dot{M}_{H}$ and the uncoupled regime occurs when the mass loss rate of hydrogen is smaller than $\sim 1.0\times10^{10}$ (or the XUV radiation is lower than 1$F_{0}$), which is consistent with the trend predicted by Equation (\ref{eq:Flux_2}). In addition, there are some differences between ours and the analytical results of \cite{Zahnle1986} and \cite{Hunten1987} (pink and blue lines) in the uncoupled regime though both results are well consistent  in the coupled regime. It should be noted that the degree of coupling is affected by other collisional processes in an atmosphere of partial ionization. Thus, we defined a general coupling factor,

\begin{equation}\label{eq:couple_factor}
 \mathscr{C}_{st} = \frac{C_{st}F_{t}}{g_rX_t}
\end{equation}
where $F_t$ is the escape flux of particles $t$, $g_r$ is the acceleration of gravity and $X_t$ is the mixing ratio of particles $t$. This parameter can be further expressed as $\mathscr{C}_{st}(r)=m_c/m_{amu}$, which defines a relative crossover mass for general collisions and demonstrates how well the minor particles $s$ coupled with main particles $t$. In the equation of \cite{Hunten1987}, the flux of helium will be cut off if $m_{c}<m_{s}$, which is similar to the behavior of $\mathscr{C}_{st}$. The regime of $\mathscr{C}_{st}<A_{s}$ means a decoupling of $s$ from $t$ . On the other hand, a large $\mathscr{C}_{st}$ reflects the tight coupling between two species. The panel (d) of Figure \ref{fig:mf_aly} shows the coupling factor $\mathscr{C}_{st}$ for some cases of group A. For the uncoupled cases (orange and red lines), most values of $\mathscr{C}_{H-He}$ of H-He collision are smaller than $A_{H_{e}}$, which means that the effect of the binary diffusion of hydrogen to helium at the bottom of the atmosphere is weak such that only minor helium can be dragged out of the gravitational well of the planet. We also notice that the $\mathscr{C}_{H^{+}-He^{+}}$ of the H$^{+}$-He$^{+}$ collision is small (orange and red dashed lines), but it is not negligible compared to $\mathscr{C}_{H-He}$. We thus further calculated the case of $\eta=0.2$ and $F_{XUV}=0.75F_{0}$ with an artificially low Coulomb collision rate ($10^{-3}$ of the original value) and found that the mass-loss rate of helium drop to almost zero as predicted by \cite{Zahnle1986} and \cite{Hunten1987}. However, one should note that for the $F_{He}/F_H$ , a value of 0.2\% predicted by our model is quite close to zero although the result of neglecting the Coulomb collision is closer. In addition, for the coupled cases (blue and purple lines of panel (d)), the coupling factors $\mathscr{C}_{H-He}$ are larger than the relative atomic mass of helium from the bottom of the atmosphere to 2 $R_{p}$, which means that the helium can be driven to high altitudes even if the Coulomb collision is neglected. In this situation, the results of \cite{Hunten1987} are almost consistent with ours. We can conclude from the above results that the coupling of particles in the lower atmosphere could be more important than those in the higher altitudes. However, we also emphasized that the conclusion should be explored with more examples.

For the case of $\eta=0.2$ and $F_{XUV}=0.60F_{0}$, the velocities of neutral helium are negative in the regions of 1.2-3.2 $R_p$ (see the panel (a3) of Figure \ref{fig:struc}) and the crossover mass calculated from our model is also smaller than the mass of atomic helium (Table \ref{tab:model_para}). As shown in panel (a) of Figure \ref{fig:he_decouple}, the escape of atomic helium seems to be separated by the regions of negative velocity. In the negative velocity region, the atomic helium can not escape because of gravity deceleration. In fact, in the lower altitudes of the atmosphere, the atomic helium can be in the state of quasi-hydrostatic equilibrium. Due to the negative velocities, no atomic helium can be transported to higher altitudes. Such behavior does not have to appear on the ions of helium. In the bottom of the atmosphere, the number of He$^+$ increases rapidly so that they can be transported to higher locations. The atomic helium can be produced by the recombination of He$^+$ and e$^-$ and charge exchange between H and He$^+$ as shown in panel (b) of Figure \ref{fig:he_decouple} so that there is a slight escape of atomic helium in the regions of higher than 3 $R_{p}$. In fact, there is a small circulation between He and He$^+$. Since the charge exchange rate between H and He$^{+}$ is greater than photoionization rate of He in the regions of 1.6-3.2$R_{p}$, some atomic helium will be produced (see panel (b) of Figure \ref{fig:he_decouple}). As a consequence of negative velocities, the helium will flow toward the planet so that the inflowed helium will be ionized again and form a mass circulation from 1.2-3.2 $R_{p}$. This inflowed He leads to an accumulation of He in the region lower than 3.2$R_{p}$, while this accumulation also enhances the ionization of helium and leads to a superfluous He$^+$ escape. As a consequence of the balance of these two antagonistic processes, the total mass-loss rate of He and He$^+$ keeps a decent conservation as shown in panel (a) of Figure \ref{fig:he_decouple}. Though $\mathscr{C}_{He+, H+}$ will grow to $10^4$ (which is fairly large compared to $A_{He^+}$) as shown in the panel (d) of Figure \ref{fig:mf_aly}, the fractionation of helium seems to be affected little by the high coupling between He$^+$ and H$^+$ in high altitudes and keeps a low value due to insufficient collision between He and H in lower altitude.

\subsection{Fitting the absorption of He 10830${\AA}$}\label{subsec:observation}
We used our results to fit the excess absorption of He in 10830 \AA. The absorption is produced by the metastable He(2$^3$S) in the atmosphere of HD 209458b. To fit the observations, we calculated the population of He(2$^3$S) and the transmission spectrum of He 10830 Åaccording to \cite{Yan2022}. The population of He(2$^3$S) was obtained by using a nonlocal thermodynamic model, which solves the rate equilibrium of He(2$^3$S). In the model, the near- and far-ultraviolet spectrum that ionizes the He(2$^3$S) population is taken from the stellar atmosphere model of \cite{Castelli2003}. When modeling the transmission spectrum, we assume that the geometry of the stellar line is in plane-parallel and that of the planetary atmosphere is in 1D spherical. Besides, the stellar He 10830 \AA \ intensity is assumed to be limb-darkened with an Eddington approximation. In fact, an one-dimensional hydrodynamic model combined with three-dimensional population calculation is self-consistent to some extent because the contribution to the He 10830 \AA \ mainly comes from the lower-middle atmospheres. In this situation, the assumption of 1D hydrodynamics causes relatively reliable results because the asymmetry of the atmosphere increases with the altitudes. For more details, please refer to the paper of \cite{Yan2022}. In order to fit the transmission spectrum of He10830, the spectral line is shifted -1.8km/s (-0.065 \AA) toward the blue side because our 1D model can not simulate the blue shift of about -1.8km/s detected by \cite{Alonso2019}.

\begin{figure}[ht!]
\plotone{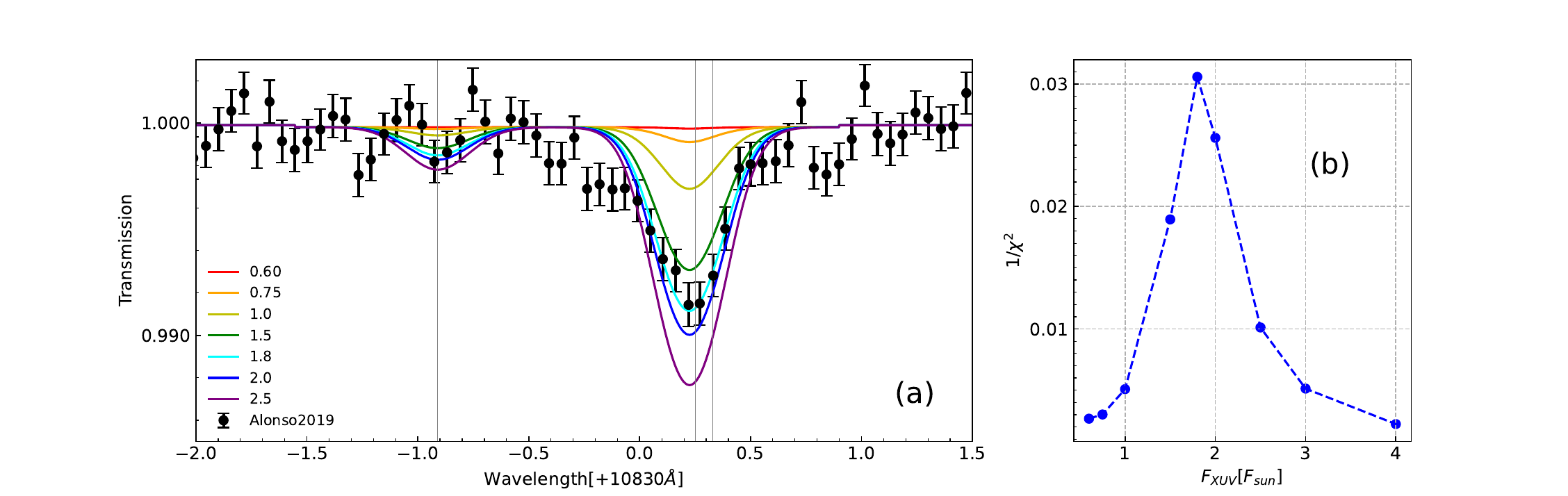}
\caption{Comparison transmission spectrum of He 10830 of the models in group A with the observation \citep{Alonso2019}. In panel (a), the transmission observation is shown by the black points with error bars. The solid lines represent the models with different $F_{XUV}$. Gray lines denote helium triplet as a comparison. Panel (b) shows $\chi^2$ of our models.}
\label{fig:fit_ob}
\end{figure}

\begin{figure}[ht!]
\plotone{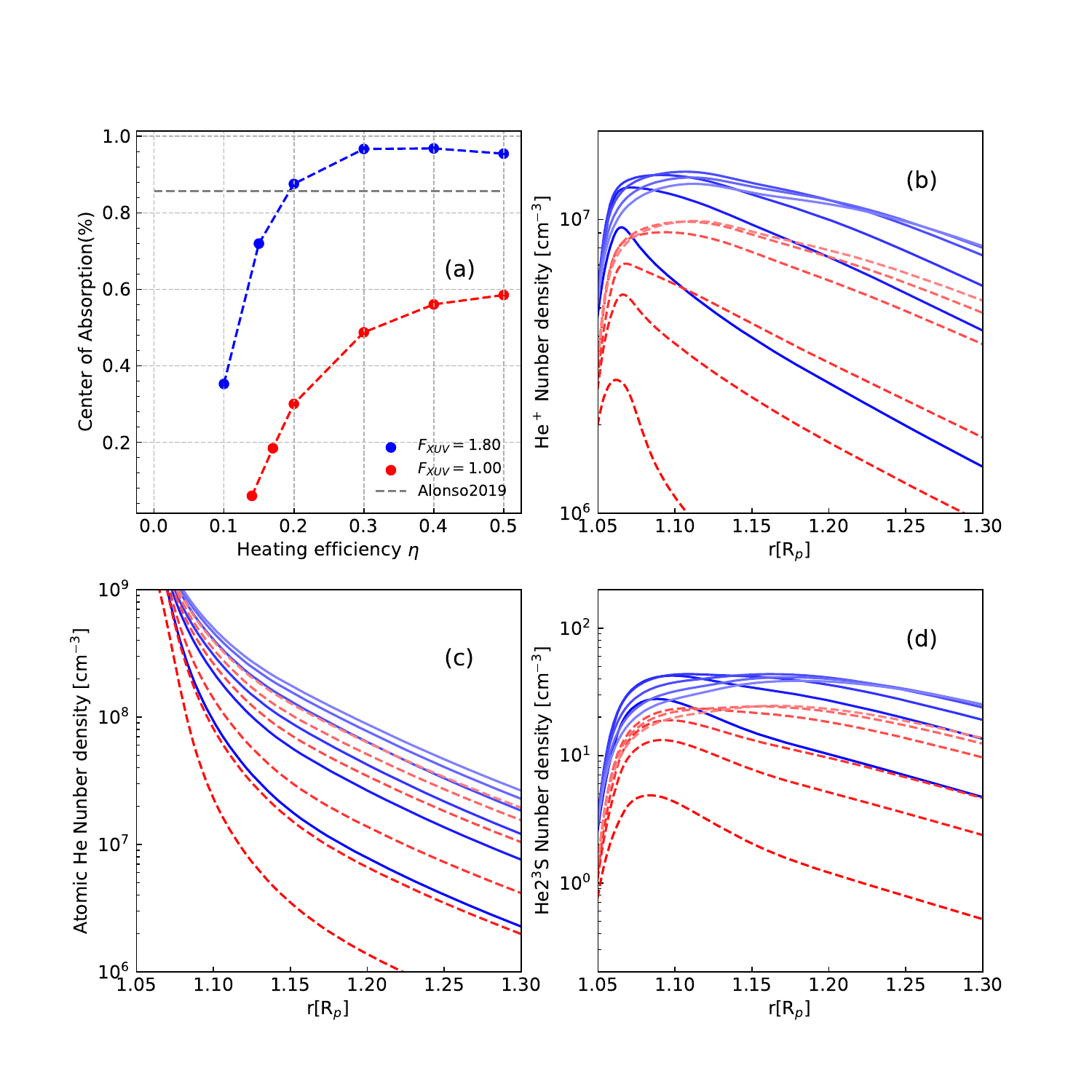}
\caption{The absorption of the line center and the distributions of the number densities of He. In panel (a) we show the absorption of line center for the models of group B (blue points) and C (red points). Gray lines represent the absorption of the observation. The number densities of He$^+$, He and He2$^3$S are shown in panels (b)-(d). The blue solid lines and red dashed lines are the number densities of group B and group C, respectively. From top to bottom, the heating efficiency is decreasing. We also set the different shades for each line.}
\label{fig:he_saturate}
\end{figure}

We compared the profile of the spectral line of our models with that of observation. We used $\chi^{2}$ method to match the observed spectral line. The best-fit parameters are determined from the results of group A-C. One can see that the best-fit case is  $\eta = 0.20$ and $F_{XUV} = 1.80 F_{0}$ (Figure \ref{fig:fit_ob}). In this case, the mass loss rate is $2.01\times 10^{10}g\ s^{-1}$ and $F_{He}/F_{H}$ is $0.039$. Though around the -13km/s (-0.47 \AA) blue shift of transfer spectrum, we didn't fit the observation signal. While in \cite{Yan2022}, the model fitting with transfer observation of Wasp-52 b didn't show this difference. The signal is hard to explain even with 3D MHD model of \cite{Khodachenko2021_0}. We thought this signal needs a more detailed physical modeling and research. Our results show that the depth of absorption increases with the increase of $F_{XUV}$ roughly linearly. However, for the case with $\eta = 0.20$ and $F_{XUV}=0.60F_{0}$ the escaping flux of helium is about $10^{-4} F_{H}$ so that the absorption depth at the line center is less than 0.001, which means that it is difficult to detect the observable signals in such atmosphere.

An increase of $\eta$ will cause more hydrogen to escape, thus also more escaping helium (see Table \ref{tab:model_para}). In this situation, our model predicts that a deeper absorption at the line center of 10830\AA \ occurs if the $F_{XUV}$ is fixed, but the heating efficiency is higher. As shown in panel (a) of Figure \ref{fig:he_saturate}, however, the absorption at the line center approaches saturation with the increase of $\eta$. The reason is that He $2^3$S is produced mainly via the recombination reaction of He$^+$ and e$^-$ \citep{Oklopcic2018}. In this situation, the production of He$^+$ is mainly controlled by the ionization of XUV irradiation because a higher heating efficiency does not increase the ionization degree but only slightly modifies the recombination rate via the variations of the temperature profiles. This explains why the number densities of He increase with the increase of $\eta$ (panel (c) of Figure \ref{fig:he_saturate}), but the number densities of H$_{e}^{+}$ and He $2^3$S are insensitive to $\eta$ in the regime of higher $\eta$ (panels (b) and (d) of Figure \ref{fig:he_saturate}).

\section{Discussions}\label{sec:Discussions}

The multi-fluid model is a powerful tool for solving the issue of fractionation. In this paper, our calculation code is based on PLUTO. Due to the low mass of the electron, solving electron equations is difficult with the multi-fluid model. In this paper, the number density and velocity of the electron are obtained from the condition of quasi-neutrality, which is quite commonly used in several multi-fluid models \citep{Toth2012, Dong2017}. But how to solve the pressure equation of the electron is a little tricky. It is different from the central difference method used in \cite{Toth2012}. Here we solved the equation of electron pressure by a split Riemann solver in which the equation of the pressure is solved solely by constructing the densities for ions and electrons (see Equations \ref{eq:c_match} and \ref{eq:c_e_mt}). Under this numerical method, the energy that belongs to the electron and the ion depends on the ratio of their pressures. If the pressure produced by the electric field is smaller than the pressure of ions, the escape of ions is driven mainly by the pressure of ions. Under this circumstance, $n'\sim n_{s}>n_{es}$. Otherwise, $n'\sim n_{es}< n_{s}$. In fact, both the corresponding pressures of $n'$ and $n_{es}$, $p_{s}$ and $p_{es}$, are included in the momentum equation (Equation \ref{eq:i_momt_3}). Thus, the specific values of $n'$ and $n_{es}$ do not affect the results. To further validate this, we calculated the tests 1-5 of \cite{Toro2009}. Our results show that the profiles of pressure, density, and temperature are independent of $n'$ (see Appendix A). In addition, to evaluate the effect of this energy distribution, we calculated an extreme case test in which kinetic energy $E_k$ is much larger than thermal energy $E_t$. As shown in Figure \ref{fig:ek_aly}, this energy distribution strategy gives a decent result compared to those calculated by PLUTO, which justifies the validation of the split method. We emphasize that our numerical method is tested using the HLL solver. Further applications in other solvers require verification.

Our models predict the phenomenon of the critical escape as discovered by \cite{Hunten1987}. A consequence of the mass fractionation is to modify the chemical composition of the planet's atmosphere. According to the best-fit cases in our models, HD 209458b could lose about 0.2\% of its mass in 10Gyr. However, this does not mean that the effect of fractionation is negligible because more helium can be retained. \cite{Hu2015} showed that a helium-dominated atmosphere caused by the fractionation of helium can explain the thermal emission spectrum of GJ 436b. In addition, the influence of the fractionation on the thermal evolution of planets should also be investigated. For exoplanets with lower mass-loss rates, this effect could be more prominent. Moreover, the fractionation of other elements can be related to the isotope ratios in the current atmosphere of Venus, Earth, and Mars \citep{Lammer2020}. Therefore, the effect of the fractionation should be explored in both the planets of our solar system and exoplanets further.


\section{Conclusions}\label{sec:Conclusions}
In this paper, we developed a self-consistent multi-fluid model for explaining the low He/H ratio required by the observation \citep{Lampon2020}. We checked the effect of our numerical method by comparing our results with tests 1-5 of \cite{Toro2009}. We found the results of the split method are completely consistent with the analytic solutions (Figure \ref{fig:org_test}). We also found that regardless of distribution of the pressure in ions and electrons, the numerical solutions match the expected analytic solutions, which justifies the correctness of our split method. Our results show that the mass fractionation of helium can produce a low He/H ratio self-consistently. Besides, this fractionation of helium is mainly controlled by the coupling in the lower atmosphere. By fitting the absorption of the spectral line in He 10830\AA \ , we found that in the condition of the heating efficiency $\eta=0.20$, the XUV flux required is 1.80 times that of the quiet Sun. In this situation, the ratio of the escaping flux of helium to hydrogen is $\sim$0.039 which is about half of the solar value. Our results mean that the fractionation in the escape of the atmosphere naturally occurs. Therefore, there is no need to decrease the helium in the atmosphere artificially. The consequence of the fractionation should be more prominent for heavier elements and need to be explored further.

\section{Acknowledgements} \label{sec:ackn}

\begin{acknowledgements}
We thank the anonymous reviewers for their constructive comments, which helped improve the manuscript. This work is supported by the Strategic Priority Research Program of the Chinese Academy of Sciences, grant No. XDB 41000000, the National Key R$\&$D Program of China (grant No. 2021YFA1600400/2021YFA1600402) and the National Natural Science Foundation of China (grants Nos. 11973082 and 12288102). The authors gratefully acknowledge the "PHOENIX Supercomputing Platform" jointly operated by the Binary Population Synthesis Group and the Stellar Astrophysics Group at Yunnan Observatories, Chinese Academy of Sciences.

\end{acknowledgements}

\begin{appendix}
\section{code verification} \label{app:code}

\begin{table}[!t]

\begin{threeparttable}[t]
\centering
\caption{Initial conditions for tests 1-5. 2-4th columns present initial condition of mass density $\rho_L$, velocity $u_L$ and total pressure of electron $p_L(total)$ and ion on the left hand side. Meanwhile, 5-7th columns show initial condition $\rho_R, u_R, p_R(total)$ in right hand side. 8-9th columns show two groups of tests with different ratio of ion pressure $p_{ion}$ and electron pressure $p_e$. The final column shows the Courant-Friedrichs-Lewy number (CFL number) which controls the integral time step of our code.}
\begin{tabular}{ccccccccccc}

\toprule
 Test& $\rho_L$ & $u_L$ & $p_L(total)$ & $\rho_R$ & $u_R$ & $p_R(total)$ & $p_{ion}$/$p_e$ & $p_{ion}$/$p_e$ & CFL &\\
  &&&&&&&(test group 1)&(test group 2)&\\

\midrule
  1& 1.0& 0.0& 1.0& 0.125& 0.0& 0.1& 5/95& 55/45& 0.05& \\
  2& 1.0& -2.0& 0.4& 1.0& 2.0& 0.4& 5/95& 55/45& 0.4&\\
  3& 1.0& 0.0& 1000.0& 1.0& 0.0& 0.01& 5/95& 55/45& 0.05&\\
  4& 1.0& 0.0& 0.01& 1.0& 0.0& 100.0& 5/95& 55/45& 0.05&\\
  5& 5.99924& 19.5975& 460.894
   & 5.99242& -6.19633& 46.0950 & 5/95 & 55/45& 0.05&\\

\bottomrule

\end{tabular}\label{tab:test_para}
\end{threeparttable}
\end{table}

\begin{figure}[ht!]
\plotone{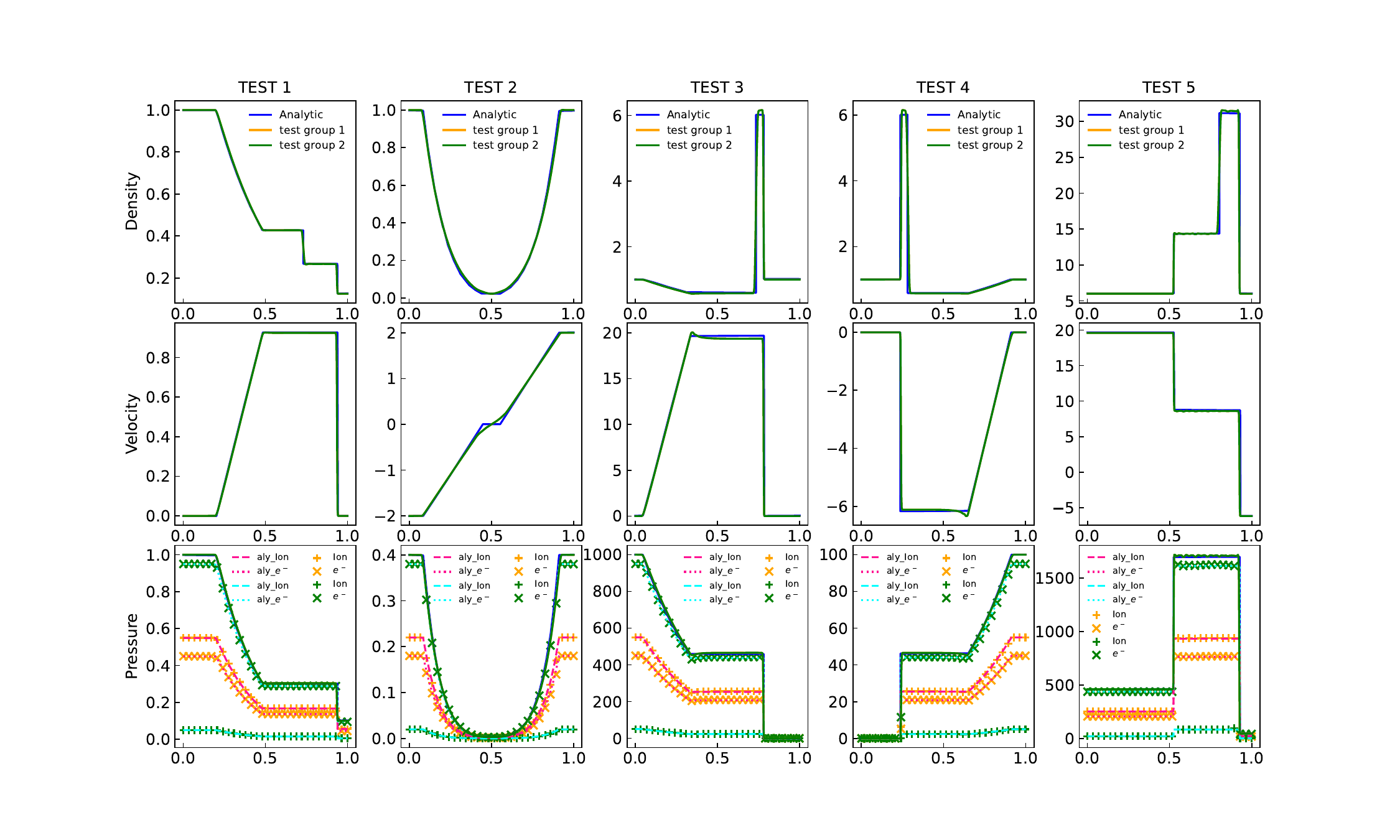}
\caption{Blue lines denote the analytic results of the tests in Table \ref{tab:test_para}. Green lines show the cases of test group 1 in Table \ref{tab:test_para}, which is an extreme case where the pressure is almost distributed to electron fluid. As a comparison, the orange lines are milder cases (test group 2 in Table \ref{tab:test_para}) in which the ratio of pressure distribution between ions and electrons is 55/45. In the pressure panels, the solid lines denote $p_{tot} = (p_s+p_e)$ of our calculation results, which is completely overlapping with that of analytic solutions. The pink dashed and dotted lines show the analytical results for ion and electron pressures when $p_{ion}/p_{e}=55/45$. Yellow + and x are the corresponding numerical results of ions and electrons for the $p_{ion}/p_{e}=55/45$ case.  The cyan dashed and dotted lines show the analytical results for the pressure of ions and electrons for the case of $p_{ion}/p_{e}=5/95$. The green + and x marks are the corresponding numerical results, respectively.}
\label{fig:org_test}
\end{figure}

\begin{figure}[ht!]
\plotone{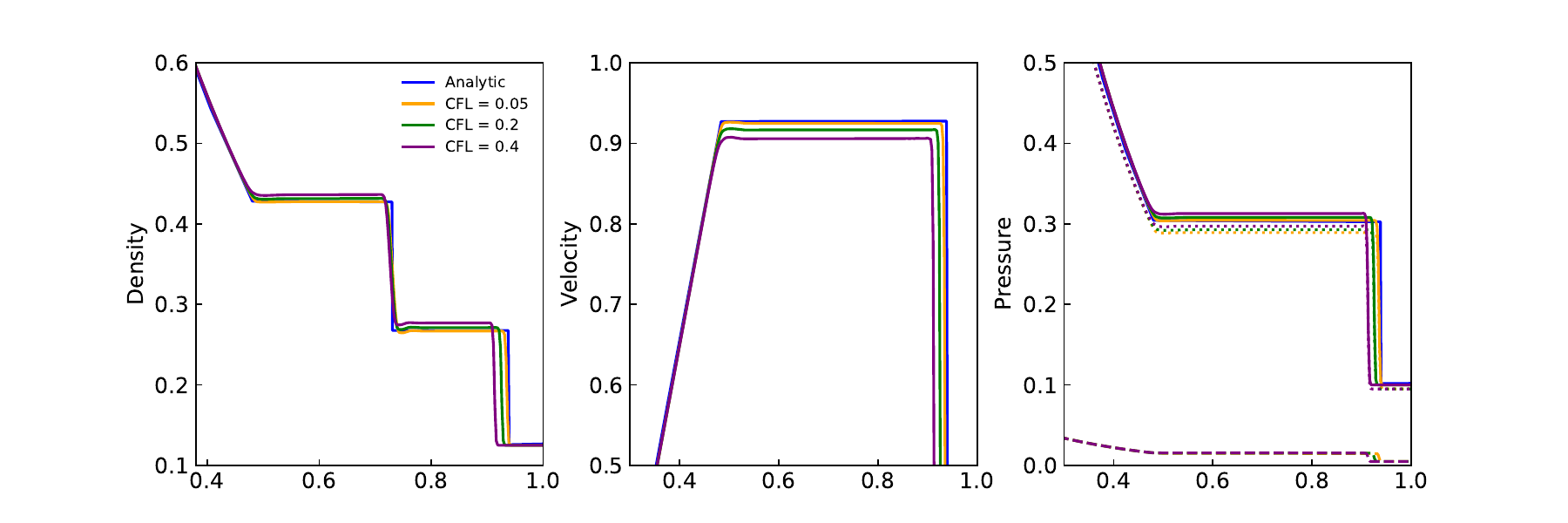}
\caption{The effect of the CFL condition on the results. The initial conditions are taken from test 1 of test group 1 in Table \ref{tab:test_para}. Same as Figure \ref{fig:org_test}, and the blue lines represent the analytical results. Yellow, green, and purple lines represent the cases of CFL = 0.05, 0.2, and 0.4.}
\label{fig:cfl_test}
\end{figure}

\begin{figure}[ht!]
\plotone{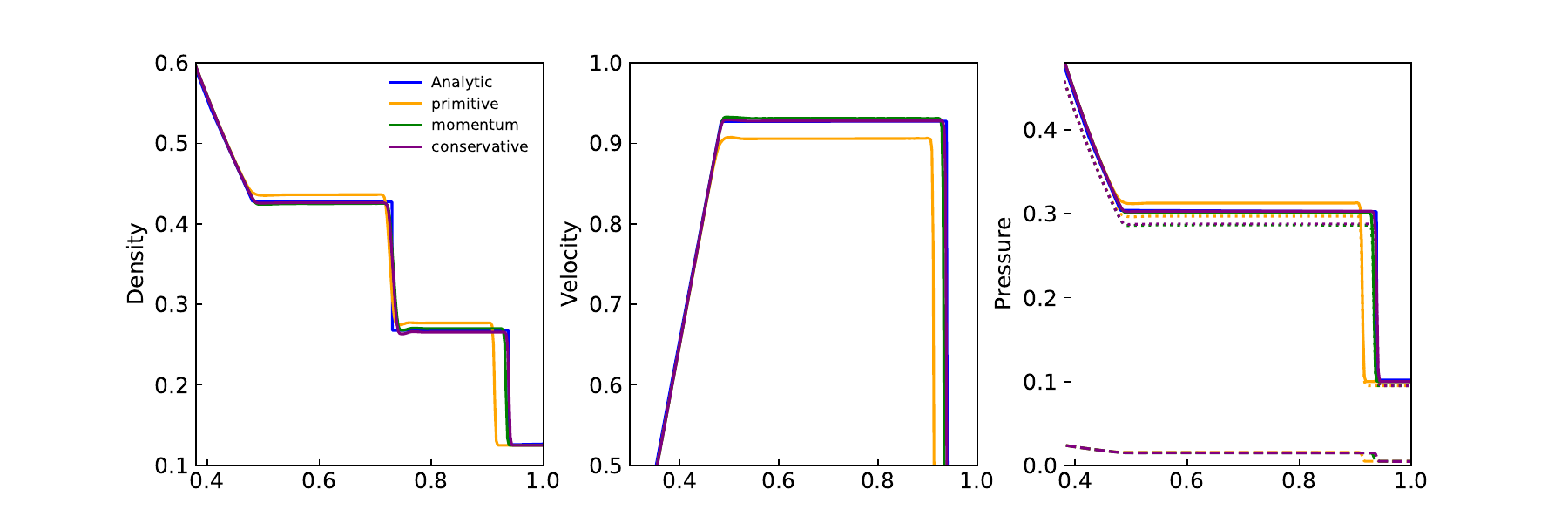}
\caption{The influence of the integrated variables on the results. The initial conditions are the same as that of Figure \ref{fig:cfl_test}. The CFL is fixed at 0.4. Blue lines are the analytical results. Yellow lines show the result when the integrated variables are density, velocity, and pressure (i.e., $\rho, u, p$). Green lines show the result when the velocity $u$ is replaced by momentum $m$ (i.e., $\rho, m, p$). The result from conservative variables (density, momentum and energy (i.e., $\rho, m, E$)) is shown by a purple line. Note that the conservative variables are also applied in origin PLUTO code.}
\label{fig:con_test}
\end{figure}

Our code is developed based on MHD code PLUTO. The validation of PLUTO has been proved \citep{Mignone2007} and its powerful capabilities were widely demonstrated in the wide field of fluid calculations. In this paper, we used a split method to describe the effect of electrons, which needs to be verified. Thus, we compared the results of the split numerical method with some benchmark plasma tests. We considered a plasma fluid composed of ions and electrons. The behavior of a fully ionized fluid is equivalent to a single fluid when the magnetic field is not included and the quasi-neutrality is valid. So we can easily compare our results with the single-fluid 1D Tests. For our calculations, the corresponding equations are

\begin{equation}\label{eq:plasma_ion_e}
\begin{matrix}
\frac{\partial{\rho}}{\partial{t}} + \nabla\cdot(\rho\textbf{u}) = 0\\
\frac{\partial{\textbf{u}}}{\partial{t}}+(\textbf{u}\cdot\nabla\textbf{u}+\frac{\nabla(p_{s}+p_{e})}{\rho})=0\\
\frac{\partial{p_s}}{\partial{t}}+[(\textbf{u}\cdot\nabla)p_{s}+\gamma p_{s}(\nabla\cdot\textbf{u}) ] = 0\\
\frac{\partial{p_e}}{\partial{t}}+[(\textbf{u}\cdot\nabla)p_{e}+\gamma p_{e}(\nabla\cdot\textbf{u}) ] = 0\\                                                                                                                                                                                                                                                                                                                                                                                   \end{matrix}
\end{equation}
where $\rho$ is the density, $\textbf{u}$ is the velocity, $p_s$ and $p_e$ are the pressure of ions and electrons, and $\gamma$ is an adiabatic index. A summation for the equations of ion and electron pressure converts Equations \ref{eq:plasma_ion_e} to the equations of single fluid when $p_{s} + p_{e}$ is defined as $p_{tot}$. In this situation, the equations can be expressed as

\begin{equation}\label{eq:plasma_single}
\begin{matrix}
\frac{\partial{\rho}}{\partial{t}} + \nabla\cdot(\rho \textbf{u}) = 0\\
\frac{\partial{\textbf{u}}}{\partial{t}}+(\textbf{u}\cdot\nabla\textbf{u}+\frac{\nabla p_{tot}}{\rho})=0\\
\frac{\partial{p_{tot}}}{\partial{t}}+[(\textbf{u}\cdot\nabla)p_{tot}+\gamma p_{tot}(\nabla\cdot\textbf{u}) ] = 0\\                                                                                                                                                                                                                                                                                                                                                                                   \end{matrix}
\end{equation}


In fact, one should notice that the properties of $p_e$ and $p_s$ in Equation \ref{eq:plasma_ion_e} are consistent with the pressure equation of Equation \ref{eq:plasma_single} because their forms are mathematically equivalent. Thus, for a given $\textbf{u}(t)$ and $\rho(t)$, one can expect that the functional forms of solution of the pressure should also be completely consistent for both cases. In addition, the behavior of $p_{tot} = (p_s+p_e)$ obtained from Equation \ref{eq:plasma_ion_e} should be accurately consistent with the pressure $p_{tot}$ of Equation \ref{eq:plasma_single}. Therefore, it can be expected that the behavior of $(p_s+p_e)$ is consistent with $p_{tot}(t)$ of Equation \ref{eq:plasma_single} and $c\cdot p_{tot}(t)$ should be the solution to the pressure equation of Equation \ref{eq:plasma_ion_e} if $p_{tot}(t)$ is the solution for the pressure equation of Equation \ref{eq:plasma_single}, where $c$ is an arbitrary constant. In fact, it is direct to expect that the $c\cdot p_{tot}(t)$ is the solution of the pressure equation of Equation \ref{eq:plasma_ion_e} because the constant c only provides a reduced factor.

In order to justify the analysis above, we compared our results with tests 1-5 of \cite{Toro2009}. The initial conditions (also see Table \ref{tab:test_para}) and the analytic solutions of $\rho$, $\textbf{u}$ and $p_{tot}$ for Equations \ref{eq:plasma_single} can be obtained in Section 4. We used 400 grid points that are distributed between 0 and 1, and all physical variables included in these tests are dimensionless. The only difference in solving Equations \ref{eq:plasma_ion_e} is that the pressures at the initial time t$_0$ in Toro's tests are distributed to ion and electron with a fixed constant c, namely, $p_s|_{t=t_0}=c\cdot p_{tot}|_{t=t_0}$ and $p_e|_{t=t_0}=(1-c)\cdot p_{tot}|_{t=t_0}$. The analytic results of these tests of Equations \ref{eq:plasma_ion_e} are not intuitive. However, one can deduce that the analytic solutions of $p_s$ and  $p_e$ in Equation \ref{eq:plasma_ion_e} are $c\cdot p_{tot}(t)$ and $(1-c)\cdot p_{tot}(t)$ owing to the same pressure equation form (.i.e, the equations for $p_{tot}$, $p_{s}$ and $p_{e}$) and sharing the same $\textbf{u}(t)$ and $\rho(t)$ in Equations \ref{eq:plasma_single} and \ref{eq:plasma_ion_e}. Thus, we preliminarily assume that the analytic solutions of pressure in Equation \ref{eq:plasma_ion_e} are $c\cdot p_{tot}(t)$ and $(1-c)\cdot p_{tot}(t)$ for $p_{s}$ and $p_{e}$, respectively. For the sake of completeness, we calculate two examples of $p_{ion}/p_e$ = 55/45 and $p_{ion}/p_e$ = 5/95, which corresponds to the cases of $p_{ion}\backsim p_e$ and $p_{ion} \ll p_e$.

Our test results are shown in Figure \ref{fig:org_test}. The numerical results justify the above analysis. First, for density, velocity, and pressure of these five tests, the results of multi-fluid (Equation \ref{eq:plasma_ion_e}) completely overlap those of single fluid (Equation \ref{eq:plasma_single}). The sum of pressure distributions of ions and electrons in Equations \ref{eq:plasma_ion_e} is accurately the same as the pressure distribution of Equations \ref{eq:plasma_single} so that it is indistinguishable in the bottom row of Figure \ref{fig:org_test}. Second, we have predicted that the analytic solutions of the pressure for ions and electrons are $c\cdot p_{tot}(t)$ and $(1-c)\cdot p_{tot}(t)$. It is clear from Figure \ref{fig:org_test} that our numerical results of $p_{s}$ and $p_{e}$ are completely consistent with the analytic results we predicted in both the mild test group ($p_{ion}/p_e$ = 55/45) and the extreme test group ($p_{ion}/p_e$ = 5/95), which justifies the correctness of our split method.

We also checked the effect of the time step on the shock tube and found that the results of smaller time steps are more consistent with the results of the analytical solution. As shown in figure \ref{fig:cfl_test}, the deviation of the shock wave is enlarged with the increase of time step, but the change of the solution in the rarefaction wave is negligible. We believe that this deviation of the shock wave is a result of using primitive integrated variables (i.e., number density $n$, velocity $u$ and pressure $p$) in our model. We compared the effect of using different integrated variables. Set 1 is for the case of primitive variable ($n, u, p$). In Set 2, we replace the $u$ by the momentum $m=\rho u$ (i.e., $n, m, p$). In Set 3, we take the full conservative variables, namely, number density $n$, momentum $m$, and energy $E=\frac{1}{2}\rho u^2 + \frac{p}{\gamma -1}$. Evidently, the results of Set 2 and Set 3 are more consistent with that of the analytic solution. This hints that the accuracy of the numerical solutions can be improved to a large extent if we use the variables of Set 2 or 3 (green and purple lines in Figure \ref{fig:con_test}). In the future, we will try to develop a code based on variables of Set 2, which can be applied to complicated HD or MHD problems, such as the interactions between stellar wind and planet wind.

\begin{figure}[ht!]
\plotone{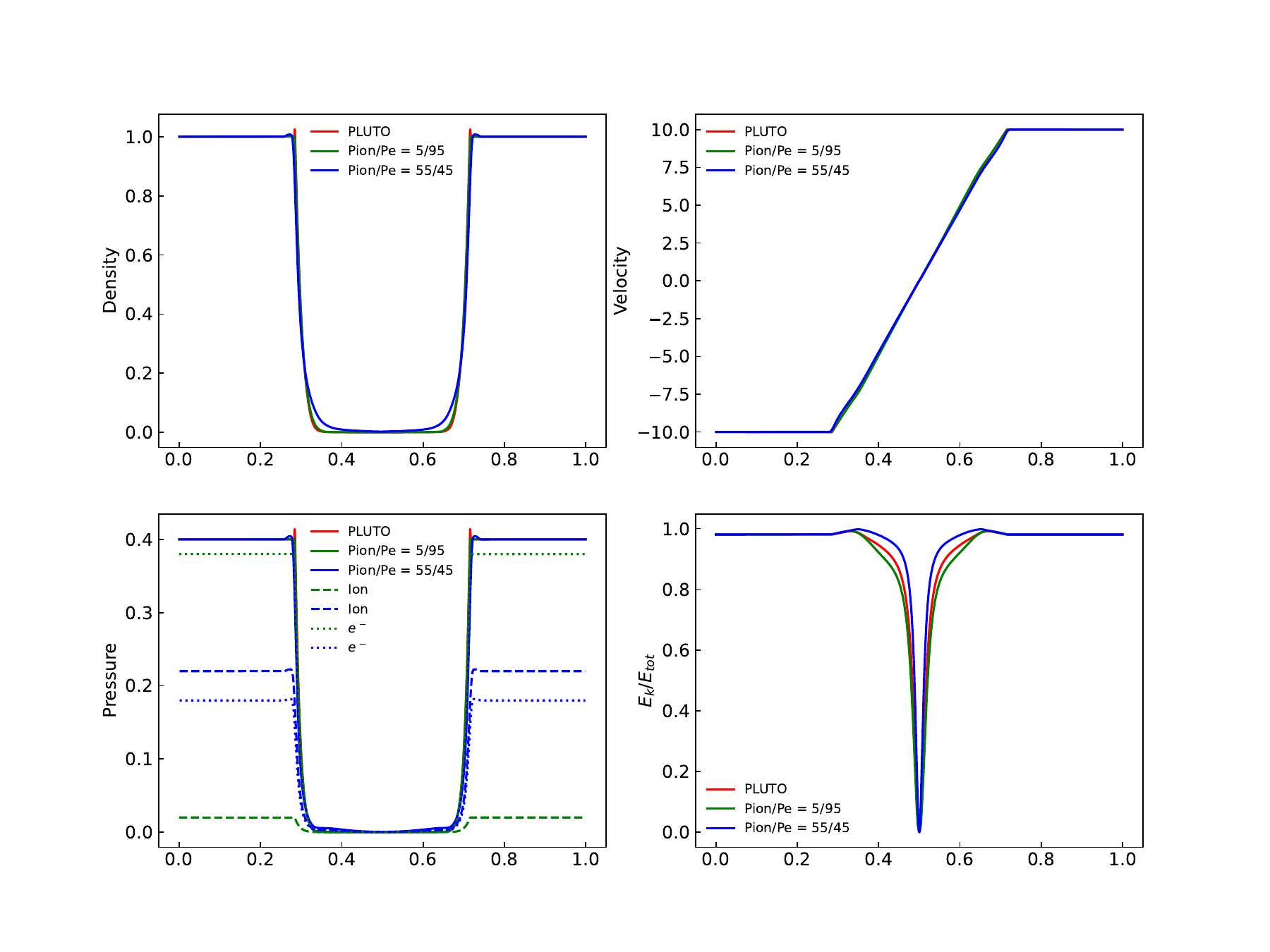}
\caption{The case with the same setting as case 2 in Table \ref{tab:test_para} while all the initial $u$ is set as 5 times the origin value. The red line is the result obtained by PLUTO with 10,000 cells. The green line is the case with $p_i/p_e$ = 5/95. The blue lines are the case with $p_i/p_e$ = 55/45 and 400 cells. The lower right panel shows the ratio of kinetic energy to total energy.}
\label{fig:ek_aly}
\end{figure}

\section{Effect of grid density in lower altitude} \label{app:grid}
\begin{figure}[ht!]
\plotone{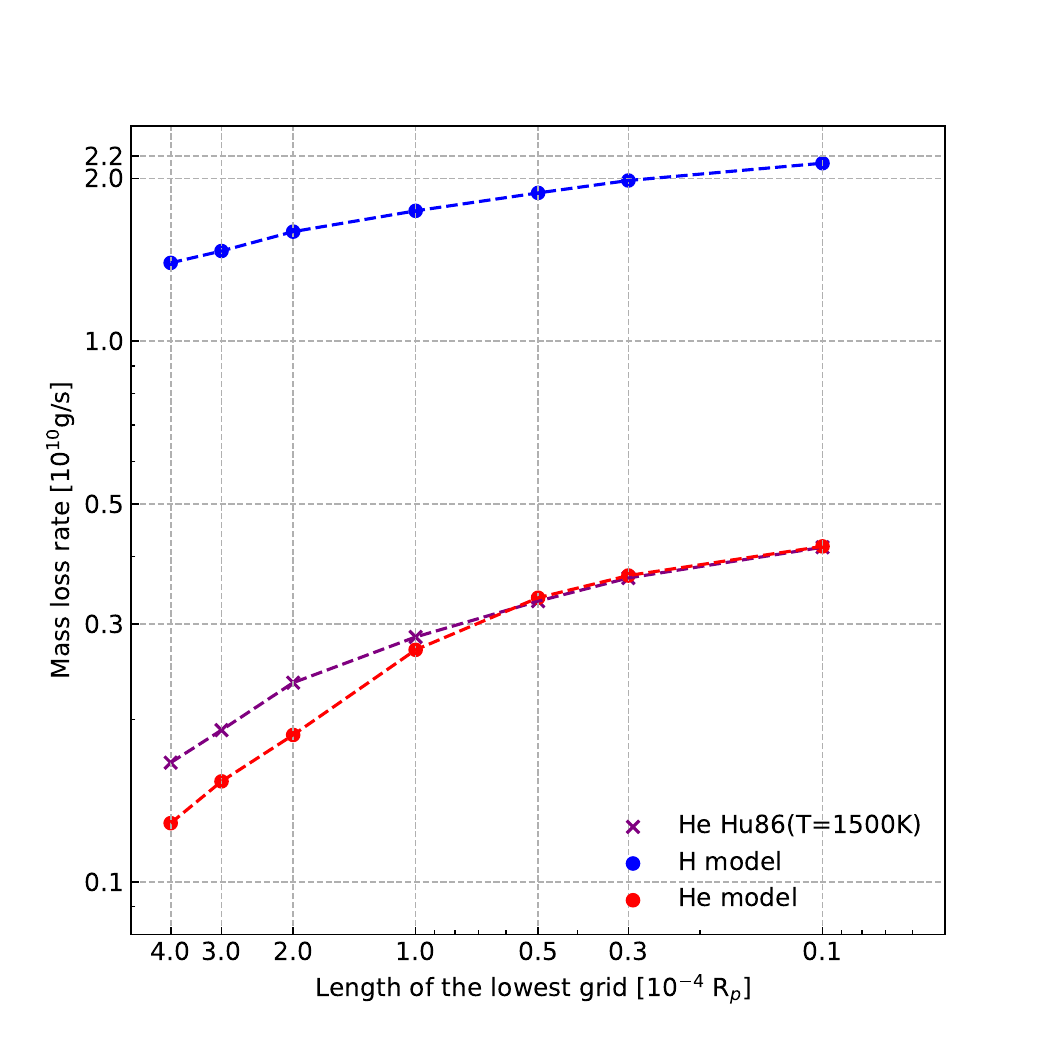}
\caption{The effect of the length of the first grid on the mass-loss rate of hydrogen and helium in the escaping atmosphere. The XUV flux is 1.8 $F_0$, and the heating efficiency $\eta$ is 0.2. The red and blue lines denote the mass loss rate of hydrogen and helium calculated by our code. The purple line represents the result of \cite{Hunten1987}.
}
\label{fig:grid_aly}
\end{figure}

In this multi-fluid simulation, the escaping rates are affected by the solution of spatial grid. A denser grid can cause a higher mass-loss rate. As shown by Figure \ref{fig:grid_aly}, the mass-loss rate can increase by as much as 50\% (from $1.39\times 10^{10}$ g/s to $2.13\times 10^{10}$ g/s) when the length of the first grid decreases a factor of 40. We have shown in the panel (c) of Figure \ref{fig:mf_aly} that there is rapid increase for the F$_{He}$/F$_{H}$ when the mass loss rates are around $\sim 1\times 10^{10}$ g/s. Therefore, the helium escaping rates increase by a factor of 3 owing to the increase in the mass-loss rates. However, the numerical resolution does not change the conclusion of this paper because there is still the fractionation of helium even if the numerical resolution is very high. We also note that in the higher numerical resolutions, our results are more consistent with the analytic result given by \cite{Hunten1987}.

\end{appendix}



\end{CJK*}
\end{document}